\documentclass[12pt]{article}
\pdfoutput=1
\usepackage{jheppub}
\usepackage{amsmath}
\usepackage{amsfonts}
\usepackage{amssymb}
\usepackage{graphicx}
\usepackage{xcolor}
\usepackage {subfigure}
\usepackage{tensor}

\usepackage[export]{adjustbox}
\newcommand{\bmat}{\left(\begin{array}}
\newcommand{\emat}{\end{array}\right)}

\def\yzero{\smash{\hbox{$y\kern-4pt\raise1pt\hbox{${}^\circ$}$}}}

\def\beq{\begin{equation}}
\def\eeq{\end{equation}}
\def\beqa{\begin{eqnarray}}
\def\eeqa{\end{eqnarray}}

\def\-{\hphantom{-}}
\def\ov{\overline}
\def\s2{\frac{1}{\sqrt2}}

\def\beq{\begin{equation}}
\def\eeq{\end{equation}}
\def\beqa{\begin{eqnarray}}
\def\eeqa{\end{eqnarray}}

\def\IF{\relax{\rm I\kern-.18em F}}
\def\II{\relax{\rm I\kern-.18em I}}

\def\Dsl{\,\raise.15ex\hbox{/}\mkern-13.5mu D} 

\def\IS{{\bf {S}}}

\def\IX{{\bf {X}}}

\def\IT{{\bf {T}}}
\def\IP{{\bf {P}}}

\def\NN{{\cal {N}}}

\catcode`\@=11   
\newdimen\@rotdimen
\newbox\@rotbox  

\def\@vspec#1{\special{ps:#1}}
\def\@rotstart#1{\@vspec{gsave currentpoint currentpoint translate
   #1 neg exch neg exch translate}}
\def\@rotfinish{\@vspec{currentpoint grestore moveto}}
\def\@rotr#1{\@rotdimen=\ht#1\advance\@rotdimen by\dp#1%
   \hbox to\@rotdimen{\hskip\ht#1\vbox to\wd#1{\@rotstart{90 rotate}%
   \box#1\vss}\hss}\@rotfinish}
\def\@rotl#1{\@rotdimen=\ht#1\advance\@rotdimen by\dp#1%
   \hbox to\@rotdimen{\vbox to\wd#1{\vskip\wd#1\@rotstart{270 rotate}%
   \box#1\vss}\hss}\@rotfinish}%
\def\@rotu#1{\@rotdimen=\ht#1\advance\@rotdimen by\dp#1%
   \hbox to\wd#1{\hskip\wd#1\vbox to\@rotdimen{\vskip\@rotdimen
   \@rotstart{-1 dup scale}\box#1\vss}\hss}\@rotfinish}%
\def\@rotf#1{\hbox to\wd#1{\hskip\wd#1\@rotstart{-1 1 scale}%
   \box#1\hss}\@rotfinish}%
\def\rotate{\@ifnextchar[{\@rotate}{\@rotate[l]}}
\def\@rotate[#1]#2{\setbox\@rotbox=\hbox{#2}\@nameuse{@rot#1}\@rotbox}

\catcode`\@=12

\begin{document}

\makeatletter
\@addtoreset{equation}{section}
\makeatother
\renewcommand{\theequation}{\thesection.\arabic{equation}}
\pagestyle{empty}
\rightline{IFT-UAM/CSIC-23-123}
\rightline{CERN-TH-2023-173}
\vspace{0.8cm}
\begin{center}
\Large{\bf Emergence of Species Scale Black Hole Horizons }
\\

\large{Jos\'e Calder\'on-Infante$^1$, Matilda Delgado$^2$, Angel M. Uranga$^2$\\[4mm]}

\footnotesize{
${}^1$ Theoretical Physics Department, CERN, \\[-0.3em]
CH-1211 Geneva 23, Switzerland}

\footnotesize{${}^2$ Instituto de F\'{\i}sica Te\'orica IFT-UAM/CSIC,\\[-0.3em] 
C/ Nicol\'as Cabrera 13-15, 
Campus de Cantoblanco, 28049 Madrid, Spain}\\ 
\footnotesize{\href{jose.calderon-infante@cern.ch}{jose.calderon-infante@cern.ch}, \href{mailto:matilda.delgado@uam.es}{matilda.delgado@uam.es},   \href{mailto:angel.uranga@csic.es}{angel.uranga@csic.es}}

\vspace*{3mm}

\small{\bf Abstract} \\
\end{center}
\begin{center}
\begin{minipage}[h]{\textwidth}

The scale at which quantum gravity becomes manifest, the species scale $\Lambda_s$, has recently been argued to take values parametrically lower than the Planck scale. We use black holes of vanishing horizon area (small black holes) in effective field theories coupled to quantum gravity to shed light on how the three different physical manifestations of the species scale $\Lambda_s$ relate to each other. (i) Near the small black hole core, a scalar field runs to infinite distance in moduli space, a regime in which the Swampland Distance Conjecture predicts a tower of exponentially light states, which lower $\Lambda_s$. (ii) We integrate out modes in the tower and generate via Emergence a set of higher derivative corrections, showing that $\Lambda_s$ is the scale at which such terms become relevant. (iii) Finally, higher derivative terms modify the black hole solution and grant it a non-zero, species scale sized stretched horizon of radius $\Lambda_s^{-1}$, showcasing the species scale as the size of the smallest possible black hole describable in the effective theory.

We present explicit 4d examples of small black holes in 4d $\NN=2$ supergravity, and the 10d example of type IIA D0-branes. The emergence of the species scale horizon for D0-branes requires a non-trivial interplay of different 8-derivative terms in type IIA and M-theory, providing a highly non-trivial check of our unified description of the different phenomena associated to the species scale.

\end{minipage}
\end{center}
\newpage
\setcounter{page}{1}
\pagestyle{plain}
\renewcommand{\thefootnote}{\arabic{footnote}}
\setcounter{footnote}{0}

\tableofcontents

\vspace*{1cm}

\newpage
\section{Introduction}

Recent activity in the Swampland Program (see \cite{Palti:2019pca,vanBeest:2021lhn,Grana:2021zvf,Agmon:2022thq} for reviews) is providing strong support that the cutoff of an effective field theory (EFT) consistent with quantum gravity is not the Planck scale, but potentially a much lower one, the species scale $\Lambda_s$ \cite{Dvali:2007hz,Dvali:2008ec,Dvali:2009ks,Dvali:2010vm,Dvali:2012uq}. This implies that objects which can be reliably described in the EFT have a lower bound in their size of order $\Lambda_s^{-1}$. This is a very profound statement since it leads to quantum gravity effects at length scales which can potentially be parametrically larger than the Planck scale. This happens for instance near infinite distance points in moduli spaces, where the (Swampland) Distance Conjecture (SDC) \cite{Ooguri:2006in} predicts the appearance of an infinite tower of light particles. Additionally, bounds on how fast these towers and the species scale become light have been proposed and tested in \cite{Etheredge:2022opl,vandeHeisteeg:2023uxj,Etheredge:2023odp,Calderon-Infante:2023ler}. 

A natural setup in which to explore and sharpen these ideas is in black hole physics. In particular, charged extremal black holes with scalar dependent gauge kinetic functions have an attractor mechanism \cite{Ferrara:1995ih,Ferrara:1996dd,Ferrara:1996um,Ferrara:1997tw} that fixes the vevs of such scalars at the horizon. By appropriately choosing the charges of the black hole, one can tune the point of moduli space that is probed by the horizon. A particular choice of charges leads to small black holes, which classically have zero horizon area (i.e. a singularity), and for which scalars run off to infinite distance as one approaches the black hole core (see \cite{Hamada:2021yxy,Angius:2022aeq,Angius:2023xtu} for other uses of small black holes to explore swampland constraints and \cite{Buratti:2018xjt,Lanza:2020qmt,Buratti:2021yia,Lanza:2021qsu,Buratti:2021fiv,Angius:2022aeq,Angius:2022mgh,Delgado:2022dkz,Angius:2023xtu,Huertas:2023syg} for other spacetime-dependent configurations probing infinite distances in field space). One would then expect that quantum gravity effects smooth out the singularity by generating a stretched horizon of small but finite size. According to our discussion above, this size should be given by the species scale $\Lambda_s^{-1}$ (hence parametrically larger than the Planck length). Indeed, it was argued in \cite{Hamada:2021yxy} that in order for small black hole solutions not to violate entropy bounds, they should acquire a finite size, of the order of the UV cut-off.

Black holes have been used at multiple occasions in the literature to probe the species scale \cite{Cribiori:2022nke,Cribiori:2023ffn}. And conversely, the species scale itself is often defined as being the size of the smallest possible black hole (SPBH) that one can reliably describe within an EFT description \cite{Dvali:2007hz,Dvali:2007wp}. One of our objectives is to study how quantum gravitational effects at the species scale come to modify small black hole solutions and ultimately promote them to these non-singular, finite-sized SPBHs, in perfect agreement with the works of \cite{Hamada:2021yxy}. 

The appearance of  stretched horizons for small black holes has long been explored in the context of string theory, e.g. for supersymmetric (and non-supersymmetric) charged black holes in 4d $\NN =4,2$ supersymmetric compactifications, by the inclusion of higher derivative corrections, resulting in string scale stretched horizons.\footnote{For a complementary viewpoint, see \cite{Cano:2018hut,Cano:2021dyy}.} Actually this corresponds in those examples to precisely the species scale in the infinite distance limit of the corresponding small black hole, in agreement with our discussion above (see \cite{Cribiori:2022nke,Cribiori:2023ffn} for related works). 

In the above discussion, the appearance of the species scale associated to the SDC tower seems accidental. We will however argue that there is a direct link, by showing that integrating out the SDC tower produces, in the spirit of Emergence \cite{Heidenreich:2017sim,Grimm:2018ohb,Heidenreich:2018kpg,Corvilain:2018lgw} (see \cite{Marchesano:2022axe,Castellano:2022bvr,Blumenhagen:2023yws,Blumenhagen:2023tev,Blumenhagen:2023xmk} for recent discussions), the higher curvature terms which ultimately lead to the stretched horizon, which is thus naturally set by the species scale. Hence, we provide an explicit microscopic mechanism for the emergence of species scale horizons for this class of small black holes.\footnote{It would be interesting to explore connections with other proposals to produce stretched horizons parametrically larger thank the Planck scale for small black holes, such as the fuzzball proposal (see \cite{Mathur:2005zp,Bena:2007kg,Chowdhury:2010ct} for reviews).} Our result moreover matches the approach in \cite{vandeHeisteeg:2022btw}, where the species scale has been argued to be the scale at which higher curvature terms become relevant. 

Hence, our work allows us to bring together three definitions of the species scale: the original definition accounting for the number of species in the SDC tower, the scale defining the size of SPBHs, and the scale of higher curvature corrections. The central idea is that small black hole solutions get resolved as SPBHs at the species scale, by including the effect of higher derivative corrections emerging from integrating out the SDC tower. Given that our main tool is small black holes, our results hold in asymptotic regimes in moduli space; it would be interesting to extend our understanding to test recent proposal involving the species scale in the interior of moduli space \cite{vandeHeisteeg:2022btw,Cribiori:2022nke,vandeHeisteeg:2023ubh}.

We expect that the emergence of species scale stretched horizons should hold beyond 4d systems. In fact, the main point of this work is to explore this question for D0-brane solutions of 10d type IIA. These behave as 10d charged small black holes, of classically zero size, and at whose core the dilaton runs off to infinitely strong coupling. Swampland considerations imply that a species scale horizon should arise. However, despite the fact that systems of $N$ D0-branes have been under scrutiny for decades from diverse perspectives, including supersymmetric quantum mechanics \cite{deWit:1988wri,Douglas:1996yp}, M(atrix) theory \cite{Banks:1996vh}, holography \cite{Itzhaki:1998dd}, their combination \cite{Ortiz:2014aja}, and MonteCarlo simulations (see \cite{Pateloudis:2022ijr} for a recent update), the only discussion about D0-brane horizons seems to have appeared in \cite{Sinha:2006yy}.

We will provide several arguments for the existence of this species scale horizon for D0-brane solutions; our arguments include microstate counting, finite temperature considerations and the analysis of higher curvature terms using (a 10d version of) the entropy function. On the other hand, we compute the familiar higher curvature $R^4$ terms arising from emergence by integrating out the SDC tower of BPS particles (which are D0-branes themselves) and show that it does not suffice to generate the horizon. Hence, the appearance of the species scale horizon implied by Swampland considerations demands the inclusion of further higher derivative corrections. Indeed, we show that one indeed gets a finite size species scale horizon once we include 8-derivative terms involving curvatures and RR 2-form field strength, a computation which we carry out using a lift to 11d. This also nicely dovetails with the fact that the species scale is the 11d Planck scale.

We regard this as impressive evidence of the non-trivial power of the species scale proposal, which provides a rationale for the presence of these extra terms (which are otherwise largely ignored in the literature), and hence probes deep UV properties of M-theory.

One general aspect in our analysis is that the limited knowledge of higher derivative corrections implies that the computations can be carried out including only the leading terms. However, the scales probed, and in particular the species scale, are such that in principle the whole set of higher derivative terms contribute in comparable amounts (hence, the use of {\em leading} is misleading). The key point however is that the main effect of the correction is to turn the singular behaviour of the two-derivative approximation into a smoothed out behaviour controlled by the species scale, and it suffices to truncate the infinite set of higher derivative terms to a tractable finite subset which captures the essence of this change, namely the presence of the species scale horizon, with the expectation that the remaining terms modify only order 1 numerical factors, but not the parametric dependence producing the species scale. In fact, in specific setups, such as 4d small black holes, this has been substantiated using scaling and other arguments \cite{Sen:1995in,Dabholkar:2012zz}. We expect this lesson to apply to other setups as well, including the 10d type IIA D0-branes. For this case, we do not have a clear argument why a truncation of the higher derivative terms suffices to capture the existence and parametric dependence of the stretched horizon. It would be interesting to argue for this, for instance by looking for general scaling properties of the higher derivative terms in the type IIA effective action with the dilaton.

\smallskip

The paper is organized as follows. In Section \ref{sec:stretched-4d} we consider small black holes in 4d $\NN=2$. Section \ref{sec:4dN2attractors} reviews the attractor mechanism, including higher curvature corrections, and in section \ref{sec:tower-4d} we study a class of 4d small black holes, the appearance of a species scale horizon from the higher curvature corrections, and the emergence of the latter from integrating out an SDC tower of KK gravitons of M-theory on $\IS^1$ (times the CY$_3$). In Section \ref{sec:stretched-10d} we turn to the case of 10d type IIA D0-branes, regarded as small black holes. In section \ref{sec:microscopic} we carry out a microscopic computation of the entropy of the system of $N$ D0-branes and of the appearance of the species scale scaling as $N^{1/2}$.  Section \ref{sec:hot} presents a complementary discussion of the computation of the entropy from the perspective of the finite temperature deformation of the D0-brane supergravity solution. 

In Section \ref{sec:final-horizon-d0} we study the appearance of the horizon from higher derivative terms in the action. In section \ref{sec:entropy-functional-d0} we perform an entropy function analysis of the spacetime solution, and show that general $R^4$ corrections can lead to the appearance of a species scale stretched horizon matching the microscopic results. In Section \ref{sec:tower-10d} we consider the  emergence of certain supersymmetric $R^4$ terms from integrating out the SDC tower in the limit explored by the solution (which is a D0-brane tower, or equivalently M-theory KK gravitons). In section \ref{sec:r4terms} we apply the entropy function to these terms and show they still do not lead to the appearance of the horizon. Finally, in section \ref{sec:mtheory-story} we show that including terms involving the RR 2-form field strength one recovers the species scale horizon for the system. We offer some final thoughts and future prospects in Section \ref{sec:conclusions}.

\section{Warmup Example: 4d Small Black Holes}
\label{sec:stretched-4d}

In this section, we illustrate our picture of the diverse appearances of the species scale using small black holes in 4d $\NN=2$ EFTs obtained from compactifications of type IIA theory on a Calabi-Yau threefold. As is well known, there are large classes of BPS black hole solutions displaying the attractor mechanism \cite{Ferrara:1995ih}: Vector multiple moduli vary along the radial direction, and, in the AdS$_2\times \IS^2$ near-horizon geometry, attain values  which are fixed in terms of the black hole charges and are independent of the asymptotic value of the moduli (while hypermultiplets remain fixed all along the flow). Small black holes correspond to solutions where the attractor mechanism drives the scalars to infinite distance in field space, leading to a horizon of formally zero size and to a singularity. As has been studied extensively in the literature (see \cite{Mohaupt:2000mj,Dabholkar:2012zz} for reviews), higher curvature corrections generically lead to a stretched horizon and a finite entropy, matching microscopic computations in many classes of examples. In section \ref{sec:4dN2attractors}, we summarize these developments and describe the case of a specific class of small black hole solutions in full detail.

Our new angle is discussed in section \ref{sec:tower-4d}, where we argue that the Swampland Distance Conjecture infinite tower of states becoming light in the infinite distance limit corresponding to small black holes is precisely responsible, via emergence, for the appearance of higher derivative terms producing the stretched horizon. We thus link the appearance of the species scale as derived from the SDC tower, as controlling higher derivative corrections, and as setting the size of the smallest possible black hole in the effective field theory. It is amusing to see that swampland ideas play such a central role in the almost three decade-long success story of microstate explanation of black hole entropy.\footnote{For other approaches to the Distance Conjecture and Black Holes, see \cite{Hamada:2021yxy,Cribiori:2022cho,Angius:2022aeq,Delgado:2022dkz,Cribiori:2022nke,Cribiori:2023swd,Angius:2023xtu}. For the interplay of higher curvature corrections and swampland constrains, mainly the Weak Gravity Conjecture \cite{ArkaniHamed:2006dz}, see e.g. \cite{Cheung:2014ega,Cheung:2018cwt,Arkani-Hamed:2021ajd}}

\subsection{4d $\NN=2$ supergravity, attractors and higher derivative corrections}\label{sec:4dN2attractors}

\subsubsection{$\NN=2$ supergravity black holes}

Type II string theory on a Calabi-Yau threefold gives 4d $\NN=2$ supergravity theories. We will focus on the physics in vector multiplet moduli space, as hypermultiplets are inert. Including the graviphoton, there are $n_V+1$ gauge bosons, labelled by $I = 0,\ldots,n_V$. The structure of the bosonic Lagrangian is
\beqa
\mathcal L = R - 2g_{i\bar\jmath}\, \partial_\mu z^i \partial^\mu \bar z^{\bar \jmath} + \hbox{Im}{\cal N}_ {IJ}\, F^I_{\mu \nu}\, F^{J\,\mu\nu} +\hbox{Re}{\cal N}_{IJ}\, F^{I}_{\mu\nu}\,\frac{\epsilon^{\mu\nu\rho \sigma}}{2\sqrt{-g}} F^J_{\rho \sigma} \, ,
\eeqa
where the different quantities are defined using special geometry (see \cite{Lauria:2020rhc} for reference). The scalars are parametrized by a set of projective coordinates $X^I$, and the K\"ahler potential is determined by the prepotential  $F(X)$, a holomorphic function of degree 2, as
\begin{equation}\label{eq:Kahlerpotdef}
    \mathcal{K}=-\log\; i \left( \bar X ^I F_ I - X^I \bar F _ I  \right)\quad ,\quad {\rm with}\; F_I=\partial_IF \, .
\end{equation}
The gauge kinetic functions ${\cal N}_{IJ}$ are also determined by special geometry, but we will skip its detailed structure.

The microscopic description in type IIA theory is as follows. There are $h_{1,1}$ vector multiplets whose gauge bosons arise from the 10d RR 3-form integrated over 2-cycles $\omega_i$, $i=1,\ldots, h_{1,1}$. The K\"ahler moduli, complexified with the integrals of the NSNS 2-form over the 2-cycles, give rise to the vector moduli; morally, the affine coordinates $Z^i$. The corresponding $F_i\equiv \partial_iF$ are asociated to the dual 4-cycles.

The structure of the prepotential in the large volume regime is
\begin{equation}
\label{eq:cubicprep}
F_0(X) = - \frac 16 C_{ijk}   \frac{X^i X^j X^k}{X^0}\,,
\end{equation} 
where  $C_{ijk}$ are the integer triple intersection numbers of the Calabi-Yau. Away from the large volume limit, the prepotential receives corrections from worldsheet instantons. Also, as discussed later, certain higher derivative corrections can also be encoded in a corrected prepotential.

The theory contains BPS black holes constructed out of D-branes wrapped on holomorphic cycles in the CY. There is a vector of electric and magnetic charges $\Gamma=(q_I,p^I)$. The central charge is
\beqa 
{\cal Z}=e^{{\cal K}/2}\left(q_I X^I-F_Ip^I\right) \, .
\eeqa
Regular black holes have a near-horizon AdS$_2\times\IS^2$ geometry.
The values of the moduli at the horizon and the entropy are fixed by the attractor equations
\beqa \label{eq:entropy4ddef}
\partial_i{\cal Z}(X^I,{\bar X}^{\bar I}, q_I,p^I)|_h=0 \quad ,\quad S=\pi|{\cal Z}(q_I,p^I)|_h \, ,
\eeqa 
where the subindex $h$ indicates the value at the horizon. The solution can be expressed as
\begin{equation}\label{eq:attractoreqs1}
    \begin{gathered}
    p^I = \text{Re}[C_h X_h ^I ]=C_h X_h ^I  + \bar C_h {\bar X_h} ^I \, , \\
    q_I = \text{Re}[C_h F_{h \,I} ]=C_h F _{h\, I}  + \bar C_h {\bar F} _{h\,I} \, ,
    \end{gathered}
\end{equation}
where we have introduced $C= - 2 i \bar {\cal Z} e^{\mathcal{K}/2}$.

In the type IIA setup, the charges $q_0$ and $q_i$ correspond\footnote{One should take into account induced D-brane charges, e.g. a single D4-brane wrapped on a K3 has $-1$ units of induced D0-brane charge due to Chern-Simons couplings to background curvature.} to D0-branes, and D2-branes  on 2-cycles, while the charges $p^0$ and $p^i$ correspond to D6-branes on the CY and D4-branes on 4-cycles.

A celebrated class (given its simple lift to M-theory \cite{Maldacena:1997de}) is obtained by considering sets of D4-branes wrapped $p_i$ times on the $i^{th}$ 4-cycle and a number $q_0$ of D0-branes, for $|q_0|\gg p^i\gg 1$. Its attractor behaviour can be easily analyzed in the large volume limit (\ref{eq:cubicprep}), see \cite{Cribiori:2022nke} for a recent application. The attractor values for moduli, and the entropy, are (we introduce $q\equiv -q_0$)
\beqa 
X^0=-\frac 12\sqrt{\frac{\frac 16 C_{ijk}p^ip^jp^k}{q}}\quad ,\quad X^i=-\frac i2 p^i\quad ,\quad S=2\pi \sqrt{\frac 16 q\, C_{ijk}p^ip^jp^k}\, .
\eeqa 
The CY volume modulus at the horizon is thus
\beqa 
{\cal V}_h=\sqrt{\frac{q^3}{\frac 16 C_{ijk}p^ip^jp^k}} \, .
\eeqa 
Note that we need $q\gg p^i$ for a reliable large volume expansion.
In section \ref{sec:tower-4d} we exploit this class of models to build small black holes and discuss its corrections.

\subsubsection{Higher derivative corrections and quantum attractors}

The discussion of higher derivative corrections to the attractor mechanism has been studied extensively in the literature. In particular, there is a class of $\NN=2$ $R^2F^{2g-2}$ F-term corrections which are computed by the topological string, and whose effects on black holes can be included systematically \cite{LopesCardoso:1998tkj,LopesCardoso:1999cv,LopesCardoso:1999fsj}, see also \cite{Mohaupt:2000mj,Pioline:2006ni} for reviews. Following these references, one uses the Weyl superfield $W_{\mu\nu}=F^+_{\mu\nu}-R^+_{\mu\nu\lambda\rho}\theta \sigma^{\lambda\rho}\theta+\ldots$, whose lowers component is the (self-dual piece of the) graviphoton, and defines $\Upsilon=W^2$. The F-term corrections can be encoded in a degree 2 generalized prepotential
\beqa
F(X^I, \Upsilon)=\sum_{g=0}^\infty F_g(X^I) \Upsilon^g \, .
\eeqa 
The lowest term $F_0$ corresponds to the usual prepotential, and higher $F_g$'s are the genus $g$ topological string amplitude. In the presence of these corrections, the attractor equations and the entropy are modified so we have
\beqa 
&& p^I=i(X_h^I-{\bar X}_h^I)\quad , \quad q_I=i(F_I(X,\Upsilon)_h-{\ov F}_I({\ov X},{\ov\Upsilon}_h)) \, , \nonumber\\
&&\Upsilon|_h=-64 \quad , \quad    S= \pi (\,|{\cal Z}_h|^2 + 4 \text{Im}(\Upsilon \partial _\Upsilon F)_h\,)\; .
\eeqa

The structure of corrections near the large volume limit is
\beqa 
F(X,\Upsilon)= - \frac 16 C_{ijk}   \frac{X^i X^j X^k}{X^0}+d_i \frac{X^i\Upsilon}{X^0}\,,
\label{corrections-r2}
\eeqa 
where the $C_{ijk}$ are the triple-intersection numbers of the Calabi-Yau and the $d_i$ are given in terms of the second Chern classes of the Calabi-Yau by
\beqa 
d_i = -\frac{1}{24}\frac{1}{64} \int_{\IX_6} c_2(T\IX_6) \wedge\omega_i\, 
\eeqa
with $\omega_i$ a basis of $H^{1,1}(\IX_6)$.
While these corrections are present for general black holes, for regular ones they are suppressed with respect to the two-derivative result. However, they are crucial for small black holes for which the classical piece vanishes, so that the extra pieces produce a non-zero entropy, namely a stretched horizon. In the following we consider a particular class of small black holes, and show that the leading correction $F_1$, which contain 4-derivative $R^2$ corrections, produce a stretched horizon. In section \ref{sec:tower-4d} we will show that these corrections precisely follow from the SDC tower and that the horizon size is controlled by the species scale.

We would like to point out that the results can also be derived in the formalism of minimization of the entropy function in an AdS$_2\times\IS^2$ ansatz near horizon geometry \cite{Sen:2005wa}. The added terms in the Lagrangian encoded in the corrected prepotential contribute to a modified entropy function. For small black holes, there is a run away but once the corrections are included, the entropy function can actually be minimized, signaling the presence of a horizon. This approach will be exploited in the 10d setup in section \ref{sec:entropy-functional-d0}.

\subsection{Small black holes, species scale horizons and the SDC tower}
\label{sec:tower-4d}

\subsubsection{An illustrative class of stretched small black holes}

We now consider a specific class of models, based on a 2-modulus version of the large volume expansion \eqref{eq:cubicprep}.  We take the prepotential
\beqa 
F=-  \frac {X^1(X^2)^2}{X^0}\,,
\eeqa 
where we have chosen $C_{122}= 6$ for simplicity. This is a simple template that can model an internal space $\IX_6$ given by K3$\times \IT^2$ (for which extensive studies of small black holes and stretched horizons have been carried out, see \cite{Dabholkar:2012zz} for a review), or K3-fibered CY threefolds. The affine coordinates $Z^1=X^1/X^0$ and $Z^2=X^2/X^0$ correspond to the sizes of the $\IT^2$ (or the base $\IP_1$ of the K3 fibration) and of the K3 (fiber), respectively.

We can easily build small black holes with a choice of non-zero charges $p\equiv p^1$, $q\equiv -q_0$, and setting $q_i,p^0,p^2=0$. The central charge, in terms of $t_i={\rm Im}Z^i$, is 
\beqa 
{\cal Z}=\frac{q+ p (t_2)^2}{2 \sqrt{2} \, (t_1)^{1/2} t_2} \, .
\eeqa 
The attractor equations imply that $t_2$ is stabilized at the horizon at
\beqa 
t_2=\sqrt{\frac qp}\,,
\eeqa 
which is still in the large volume regime if $q\gg p$, as we assume in the following. The attractor mechanism also leads to
$t_1\to \infty$, so this scalar runs off to infinite distance in moduli space. Consequently the overall volume in string units at the horizon also diverges:
\beqa \label{eq:volumeblowsup} e^{-\mathcal{K}_h} |X_h^0|^{-2} = 8 \mathcal{V}_h = 8 t_1 t_2^2 \to \infty\,,
\eeqa
where the $|X_h^0|^{-2}$ factor accounts for the choice of projective coordinates on the special K\"ahler manifold such that $\cal V$ is independent of the K\"ahler gauge. Using the formulas \eqref{eq:entropy4ddef}, the area of the horizon and therefore the entropy vanish:
\beqa
S= \lim_{\substack{t_1\to \,\infty \\ t_2= \sqrt{\frac qp}}}\pi |{\cal Z}| \to 0\,.\eeqa

We now show that the small black hole acquires a stretched horizon upon the inclusion of the correction of the kind \eqref{corrections-r2}. In particular, we focus on the correction $d_1$, so the prepotential is
\beqa 
F=-\frac {X^1(X^2)^2}{X^0} + d_1\frac{X^1\Upsilon}{X^0} \, ,
\label{corrected-prepo-example}
\eeqa 
where the last two terms correspond to $F_1$, the genus 1 topological string amplitude. This corresponds to the class of  solutions considered in \cite{Cribiori:2022nke} with $p^i =0, \; i \neq 1$ (see also \cite{Dabholkar:2012zz} for a related class).

The moduli at the horizon and entropy are given by
\beqa \label{eq:4d-entropy}
Z^1= i \frac{\sqrt{p\, q}}{2 \sqrt{d_1 \Upsilon_h }}  \quad , \quad Z^2=0  \quad , \quad S\sim 4 \pi  \sqrt{d_1 \Upsilon_h\, p \, q}\,.
\eeqa 
We see that the horizon no longer explores an infinite distance in moduli space and that the entropy is non-vanishing. Hence the solution develops a stretched horizon that cloaks up the singularity. The overall volume at the stretched horizon is modified by the curvature corrections (see \cite{Mohaupt:2000mj} and references therein) and is given by: 
\beqa e^{-\mathcal{K}_h} |X_h^0|^{-2} = 8 \mathcal{V}_h \sim \frac{q (t_1)_h  \, }{p} \sim \frac{q^{3/2}}{\sqrt{p \, d_1 \Upsilon_h}} \, , \eeqa
which is finite, and still in the large volume regime if $q\gg p$.

Recall that the above result could have been obtained using the entropy function formalism, but we skip its explicit discussion. Rather we turn to the discussion of the appearance of the higher derivative correction from the SDC tower.

\subsubsection{The species scale horizon}

In a $d$-dimensional EFT weakly coupled to Einstein gravity, the species scale is given by:
\begin{equation} \label{eq:speciesscaledef}
    \Lambda_{s} \sim \frac{M_{d}}{N_{s}^{\frac{1}{d-2}}}\,,
\end{equation} where $N_s$ is the number of species below the species scale and $M_{d}$ is the $d$-dimensional Planck mass. In EFTs obtained from string theory (or compactifications thereof), the number of light species changes as one moves along the moduli space. Thus, the species scale is dependent on the region of moduli-space under consideration. In particular, in infinite distance limits in moduli space where infinite towers of states are becoming exponentially light \cite{Ooguri:2006in}, the species scale falls as the number of light species increases drastically. In order to determine the species scale, we must therefore study what infinite distance limit the (formerly) small black hole was probing.

Recall that the small black holes described in the previous section probes the infinite distance limit $t_1 \to \infty$, with $t_2$ fixed. Note that in this case, the overall volume modulus \eqref{eq:volumeblowsup} ${\cal V}\sim t_1 (t_2)^2$ also goes to infinity at the horizon, where ${\cal V}$ is the volume of the CY $\IX_6$ in string units. At fixed 4d Planck scale $M_p^2=M_s^2 \mathcal{V}/g_s^2$, the large ${\cal V}$ limit corresponds to a large $g_s$ limit in string variables (a similar conclusion follows from the fact that the 4d dilaton lies in a hypermultiplet, so it is constant along the attractor flow, hence $\mathcal{V}/g_s^2$ remains fixed). Hence, the limit of our interest correspond to a large $g_s$ regime. 
There is therefore a tower of D0-brane particles becoming light; this also follows from the fact that the central charge for $k$ D0-branes is given by $Z_{\rm D0}=k (t_1)^{-1/2} (t_2)^{-1}$, which goes to zero in the limit. Actually, there may be other towers of particles becoming light, but they correspond to branes wrapped on some internal cycles, so they are suppressed in the regime $q\gg p$. The D0-brane particles thus encode the leading correction. Incidentally, we also note that the infinite distance limit explored by the singular two-derivative solution is a standard decompactification limit to M-theory compactified on $\IX_6$, rather than an emergent string limit \cite{Lee:2019xtm,Lee:2019wij}. Indeed, one can show that the volume of the Calabi-Yau in M-theory units is related to the 4d dilaton and is therefore constant, signalling that there is no decompactification to 11d M-theory.

The computation of the species scale in this infinite distance limit can be done in different ways. 
We choose to focus on the approach of \cite{vandeHeisteeg:2023ubh}, where it was argued that higher derivative corrections to the EFT could be used as a proxy for determining the species scale. In this framework, the species scale acts as the energy scale that suppresses higher derivative corrections to the EFT. This recipe was applied to the case of Calabi-Yau compactifications of type II string theories to find that the species scales is given by:
\begin{equation}
    \Lambda_{s} = \frac{M_{4}}{\sqrt{F_1}}\,,
\end{equation}
where once more, $F_1$ is the genus 1 topological free energy. Taking \eqref{eq:4d-entropy} and $F_1 \sim Z^1 d_ 1 \Upsilon$, we see that at the horizon of the stretched black hole, we have: 
\begin{equation}
    S \sim \left(\frac{\Lambda_{s}}{M_{4}}\right)^{-2}\,.
\end{equation}
The radius of the stretched horizon is therefore given exactly by the species scale. 

Evaluating the species scale can also be done by counting the number of D0 states that are light in the EFT, in the spirit of \cite{Marchesano:2022axe,Castellano:2022bvr,Blumenhagen:2023yws,Blumenhagen:2023tev,Blumenhagen:2023xmk}, using \eqref{eq:speciesscaledef}. In the next section we will bridge these two interpretations of the species scale by showing how the higher derivative corrections to the EFT Lagrangian originate from integrating out the tower of D0 states, in the spirit of Emergence.

\subsubsection{The SDC tower and Higher Derivative Emergence}
\label{sec:1loop-4d}

The SDC implies that the description of infinite distance limits in moduli space require the inclusion of microscopic physics, in particular an infinite tower of states whose masses decrease exponentially with the field distance \cite{Ooguri:2006in}. On the other hand, in the previous section we have argued that the singular behaviour of the solution is cloaked by a stretched horizon if suitable higher derivative corrections are included. In this section we show that the two UV ingredients are related, and in fact the distance conjecture {\em implies} the appearance of higher derivative corrections of the necessary kind to generate the stretched horizon. Hence, the two explanations are two sides of the same phenomenon and link the different physical interpretations of the species scale.

The key idea is that the $R^2$ terms can be seen as generated by an infinite tower of 4d BPS D0-brane states, namely KK gravitons running on an $\IS^1$ compactification of the 5d theory given by M-theory on $\IX_6$. In fact, they correspond to the leading term in the Gopakumar-Vafa expansion \cite{Gopakumar:1998ii,Gopakumar:1998jq} in the large volume limit (where the contribution from D2-branes is subleading).

Indeed, M-theory on $\IX_6$ contains 5d BPS states which correspond to 11d gravitons on the groundstate of the $\IX_6$ compactification. These are given by the K3 cohomology classes, giving an overall multiplicity of 24, times a linear dependence on the K\"ahler modulus of $Z^1$. These 5d BPS states can run with KK momentum in the $\IS^1$ compactification to 4d, leading to the 4d tower of D0-brane states.\footnote{The D0-brane tower is actually analogous to that we will consider in the 10d context in section \ref{sec:tower-10d}, as they are simply related by compactification.}

Hence the problem is just the computation of 1-loop diagram, see \cite{Green:1997as} for computations in this spirit (see also \cite{Ooguri:1996me,Gutperle:1999dx,Collinucci:2009nv,Petersson:2010qu,Piazzalunga:2014waa,Gonzalo:2018guu} for related computations in other setups) . For later convenience, we present the general $n$ graviton scattering amplitude in $d$ dimensions in the worldline formalism:
\begin{equation}
\mathcal{A}_{d,n}=\frac{1}{n!\,\pi^{d/2}} \sum_{k} \int d^d\mathbf{p} \int_{0}^{\infty} \frac{d\tau}{\tau} ~\tau^n ~e^{-\tau \left( \mathbf{p}^2 + \frac{k^2}{R^2}  \right)}~.
\label{general-amp}
\end{equation}
Here $\tau$ is the worldline parameter, $k$ is the KK momentum, and $R$ is the $\IS^1$ radius.

We want to obtain the $R^2$ corrections to the effective action, which are encoded in two graviton scattering. For pedagogical reasons, it is easier to start from the 6d perspective of M-theory on K3$\times \IS^1$ and postpone compactifying on the $\IT^2$. Thus, we particularize to $d=6$, $n=2$. The above expression corresponds to the contribution of a single K3 ground state, and diverges in two ways: because of the integral and because of the infinite sum. Since the sum over momentum modes is infinite, we can perform a Poisson resummation, so that both are nicely combined. We get
\begin{equation}
    \frac{1}{\pi^{3/2}}{\cal A}_{6,2} = \pi^{3/2} \tilde{K} \int_{0}^{\infty} d\hat{\tau} \, \hat{\tau}^{1/2} \sum_{l} e^{- \pi \hat{\tau} R^{2} l^2} \, = C \tilde{K} + \frac{\zeta(3)}{\pi R^3} \tilde{K} \, ,
\end{equation}
where $\hat \tau = \tau^{-1}$. This Poisson summation trades the KK momentum $k$ for the winding number $l$ of the worldline along the $\IS^1$. The only divergence is now in the $l=0$ piece, which has been isolated as the first terms in the second equality; $C$ is an  unknown coefficient regularizing the divergence. Including the K3 ground state multiplicity of 24 and the $\IT^2$ modulus dependence upon compactification to 4d, we have 
\begin{equation}
    \frac{1}{\pi^{3/2}}{\cal A}_{4,2}   = 24 Z^1 \left[\, C \tilde{K} + \frac{\zeta(3)}{\pi R^3} \tilde{K} \,\right] \, ,
    \label{eq:amplitude-result-4d}
\end{equation}
where ${\tilde K}$ is a 4d kinematical matrix. This is of the form introduced in the prepotential \eqref{corrected-prepo-example}, hence it is precisely of the form required to produce a stretched horizon.  We note that the same 4d result (\ref{eq:amplitude-result-4d}) could have been obtained via a 1-loop computation in the 10d theory compactified directly on K3$\times \IT^2$, with the $Z^1$ prefactor arising from the momentum integral over the $\IT^2$. Hence, the 4d coupling of interest arises from higher derivative emergence.

This nicely illustrates the links among the appearance of the species scale from the SDC tower, the higher-curvature corrections and the stretching of a small black hole into the SPBH.

We finish with a comment in hindsight. It is a well-known fact that the $R^2$ corrections encoded in the corrected prepotential can be seen to arise from the compactification of suitable 10d $R^4$ terms \cite{Antoniadis:1997eg}. For concreteness, focusing on the particularly simple case of compactification on $\IX_6=$K3$\times \IT^2$, we start from the familiar 10d  supersymmetric $R^4$ terms, and put two curvature insertions in K3 to get a 6d $R^2$ term, with a prefactor of  $\chi(K3)=24$; upon further integration over $\IT^2$ (which provides no background curvature) we obtain the 4d $R^2$ term with the $Z^1$ prefactor. It is easy to generalize this argument to general CY$_3$ compactifications and obtain the corrections\footnote{Actually, there is one further constant correction to the prepotential, proportional to the Euler characteristic $\chi(\IX_6)$, arising from integrating $R^3$ over the space $\IX_6$, and reabsorbing the resulting correction to the 4d Einstein term via a change of frame \cite{Antoniadis:1997eg}. This will not be relevant for our discussion.} \eqref{corrections-r2} with $d_i$ essentially determined by the second Chern classes of $\IX_6$
\beqa 
d_i\sim c_{2\,i}=\int_{\IX_6} R\wedge R \wedge \omega_i \, ,
\eeqa 
where $\omega_i$ are the Poincar\'e dual 2-forms of the basis 2-cycles. In fact, the above is the simplest way to derive the 4d $R^2$ corrections \cite{Antoniadis:1997eg} (see \cite{Cribiori:2022nke} for a recent application). The dimensional reduction just described underlies the fact that the 4d $R^2$ terms can be obtained from integrating out a tower of 5d KK gravitons is related to a similar derivation of 10d $R^4$ terms from 11d M-theory gravitons, as we describe in section \ref{sec:tower-10d}.

\section{D0-branes and the Species Scale Horizon}
\label{sec:stretched-10d}

We now turn to the discussion of the system of $N$ D0-branes in 10d type IIA theory. We will argue for the existence of a stretched horizon controlled by the species scale, and explore its appearance from higher derivative couplings and their interplay with emergence upon integrating out the states in the corresponding SDC tower.

We start by describing D0-brane systems as small black holes in the theory.
The relevant part of the string-frame type IIA  action is
\beqa 
S= \frac{1}{2\kappa ^2} \int d^{10} x \sqrt{-g} \Big[e^{-2 \phi}\left( R + 4 \partial_\mu \phi \partial ^\mu \phi \right) - \frac{1}{2}|F_2|^2 \Big]\,,
\eeqa
where $ 2 \kappa^2 = ( 2 \pi)^7 \alpha'^4$.
The D0-brane solution is
\beqa 
   &&  ds^2 = f(r) ^{-1/2}(- dt^2) + f(r)^{1/2}(dx_{\perp}^2)\nonumber \, ,\\
  &&  e^{( \phi - \phi_\infty)}= f(r)^{3/4}\, ,\\
   && f(r) = 1+ \frac{\rho^7  g^\infty_s N}{r^7},\;\;\;\; {\rm with}\;\rho^7=(4\pi)^{\frac 52}\,\alpha'^{\frac 72} \,\Gamma\bigg( \frac{7}{2}\bigg)\nonumber \, .
    \label{d0-string-frame}
\eeqa 
We will be more interested in the Einstein frame action
\begin{equation}
  S_{\text{IIA}} = \frac{1}{2} \int d^{10}x \sqrt{-g} \, \left( R - \frac{1}{2} (\partial \phi)^{2} - \frac{1}{2} e^{\frac{3}{2} \phi } |F_2|^2 \right) \, ,
\end{equation}
where, from now on, we use 10d Planck units.
The D0-brane solution reads
\beqa \label{eq:D0einstein}
&&  ds^2 = f(r) ^{-7/8}(- dt^2) + f(r)^{1/8}(dx_{\perp}^2)\nonumber \, ,\\
  &&  e^{( \phi - \phi_\infty)}= f(r)^{3/4}\, ,\\
   && f(r) = 1+ \frac{\rho^7 g^\infty_s N}{r^7},\;\;\;\; {\rm with}\;\rho^7=(4\pi)^{\frac 52}\,\alpha'^{\frac 72} \,\Gamma\bigg( \frac{7}{2}\bigg)\nonumber \,.
\label{sugra-d0-sol}
\eeqa  

Although not usually described from this viewpoint, this solution provides an interesting analogy with small black holes in lower dimensional theories. It has an $\IS^8$ horizon of zero size at its core, where a scalar (the dilaton) goes off to infinite distance in moduli space, resulting in a singularity. The latter is not worrisome, because it simply reflects the presence of a source, whose microscopic description is beyond the two-derivative supergravity approximation.

On the other hand, swampland arguments would imply that, as we are driven to the core of the solution, namely near infinite distance limit in the dilaton moduli space, the effective field theory should include additional degrees of freedom, corresponding to an SDC tower of states. Inclusion of these in the effective field theory leads to a lowering of the cutoff scale and hence a minimal size for the system, which would naturally be associated to a species scale horizon. In the following sections we provide several arguments that such a horizon exists and is given by the species scale. 

\subsection{Microscopic derivation of the entropy and of the species scale}
\label{sec:microscopic}

The most direct way to argue that the D0-brane system develops a finite size horizon is to show that it has a large number of microstates, to which one can associate an entropy. This provides a lower bound on the effective size of the system, by holographic bounds.\footnote{Note that in general the Bekenstein-Hawking area law gets corrected when the action includes higher curvature terms; hence, this relation will be made more precise in section \ref{sec:entropy-functional-d0}.} In this section we perform the microscopic computation of the entropy of a system of D0-branes with total charge $N$ (which is familiar from similar combinatorics problems in string theory), and also explain the microscopic derivation of the species scale (which is a novel result). 

Recall that consistency of the description of 10d type IIA as M-theory on $\IS^1$ requires that, for each value of $k$, there is exactly one threshold bound state in the $k$ D0-brane system (corresponding to the 11d graviton with $k$ units of KK momentum) \cite{Witten:1995ex}. Hence, for a total D0-brane charge $N$, the number of microstates is given by the number of partitions $p(N)$.
The asymptotic behaviour for large $N$ is given by the celebrated Hardy-Ramanujan formula
\begin{equation}
    p(N) \sim \frac{1}{4\sqrt{3}\, N} \exp\left( \pi \sqrt{\frac{2}{3}} \, \sqrt{N} \right) \, .
\end{equation}
The entropy of this thermodynamic ensemble for large charge is thus
\begin{equation} \label{eq:D0-entropy}
    S \sim \sqrt{N} \, ,
\end{equation}
up to $\log N$ corrections. It is tantalizing to speculate the latter may be related to similar log corrections to the species scale discussed in \cite{Cribiori:2023sch}.

The above expression implies that the corresponding horizon size in a gravitational description of the system is parametrically larger than the Planck scale, as we will explain in later sections. In fact, the expectation that it corresponds to the species scale can already be supported from the viewpoint of microscopic combinatorics, as we show next.

One may have expected that, in order for a charge $N$ black hole to be describable in an effective field theory, the latter should include at least $N$ species (i.e. different D0-brane bound states). However, this is not correct and highly overestimates the number of species. In fact, rephrasing the definition in \cite{vandeHeisteeg:2023ubh}, the number of species $N_s$ should be the minimum necessary to explain the entropy of  black hole entropy. In other words,\footnote{Similar counting problems arise in many other setups, for instance characterizing 1/2 BPS states with potentially large R-charges in AdS/CFT, see e.g. \cite{Berenstein:2004kk,Balasubramanian:2005mg}.} if we consider a theory with the number of species given by $k$ and want to build a charge $N$ black hole, the number of ways to do so is given by $p_k(N)$, the number of partitions of $N$ with each part {\rm not larger} than $k$. This is equivalent to the number of partitions of $N$ into at most $k$ parts.\footnote{The equivalence is clear by using the relation of partitions with Young diagrams. A partition of $N=l_1+\ldots +l_k$ with $l_i\geq l_{i+1}$ with {\em at most} $k$ parts can be depicted as a Young diagram with $N$ boxes arranged in $k$ rows of lengths $l_i$. By conjugating the diagram, i.e. exchanging the roles of rows and columns, it corresponds to a partition of $N$ with arbitrary number of parts, but each part being {\em at most} $k$.} A classic result in \cite{erdos-lehner} gives the asymptotic behaviour
\beqa \label{eq:pkN}
{\rm lim}_{N\to \infty}\frac{p_k(N)}{p(N)}=\exp\bigg(-\frac 2C e^{-\frac 12 Cx}\bigg) 
\eeqa 
for 
\beqa \label{eq:def-k}
k=C^{-1}N^{\frac 12 }\log N+xN^{\frac 12}\; ,\quad {\rm with}\;\; C=\pi \sqrt{\frac 23}\, .
\eeqa 
Summing up, on the one hand the entropy of a black hole made out of $N$ D0 branes in a theory with $k$ species is given by $p_k(N)$ in \eqref{eq:pkN}. On the other, we know the entropy of such a black hole is given by $p(N) \sim \sqrt{N}$ by microstate counting. We now show that this scaling can be also recovered if and only if the number of species in the theory is at least of order $\sqrt{N}$. Indeed, equation \eqref{eq:pkN} gives an order one number as $N\to\infty$ (so that both $p(N)$ and $p_k(N)$ are of the same order) unless $x\to -\infty$. Imposing that $x$ should be bounded from below, \eqref{eq:def-k} automatically implies that $k \gtrsim \sqrt{N}$ asymptotically, where we are ignoring a multiplicative $\log N$ correction. Hence, the number of species $N_s=\sqrt{N}$ suffices to explain the black hole entropy scaling. Equivalently, most microscopic configurations can be built using blocks of bound states of at most $N_s=\sqrt{N}$ D0-branes. 

In fact, we incidentally note that the distribution is dominated by states containing at least one bound state of {\em exactly} $N_s=\sqrt{N}$ D0-branes. Following \cite{crandall}, we introduce the number $p(N,k)$ as the number of partitions of {\em exactly} $k$ parts (equivalently, partitions with at least one part {\em exactly equal} to $k$). One can define a probability measure for the likelihood of a random microstate containing at least one bound state of $k$ D0-branes:
\beqa
f_{N,k}=\frac{p(N,k)}{p(N)}\; ,\quad \sum_{k=1}^N f_{N,k}=1\,.
\eeqa
Defining the quantity
\beqa
X(k)=\frac{k}{\sqrt{N}}-\frac 1C \log n\; ,
\eeqa
the probability distribution takes the asymptotic form
\beqa
f_{N,k}\sim e^{-\frac C2 e^{-\frac 2C X(k)}}-e^{-\frac C2 e^{-\frac 2C X(k-1)}}\,.
\eeqa
The probability was shown in \cite{skezeres} to be maximized at a value
\beqa
k_0\sim\frac{\sqrt{6}}{\pi}\sqrt{N}L+\frac{6}{\pi^2}(3(L+1)/2-L^2/4)-\frac 12  \; ,\quad {\rm with}\;\;L=\log (\sqrt{6N}/\pi) \, .
\eeqa
This shows that the species scale is exactly given by $N_s=\sqrt{N}$ (and for instance, not smaller), as we had anticipated.

In the previous description, we simply used the microscopic D0-brane combinatorics, and the abstract definition of species scale. In an actual spacetime description of the system, it will be manifest that the D0-brane stack is a small black hole exploring the infinite distance limit of strong coupling, i.e. decompactification to 11d Mtheory. This will allow for the interpretation of the species scales as the 11d Planck scale. We thus turn to the description of the stretched horizon and the species scale from the spacetime perspective.

\subsection{Hot D0-branes}
\label{sec:hot}

In the previous section we have argued that the D0-branes solution develops an species scale-sized stretched horizon.
Naively, a direct approach to confirm the existence of a stretched horizon would be to solve the equations of motion including all the higher derivative corrections. However, given the lack of knowledge about the precise form of all such corrections, this is actually not feasible. In this section, we propose an alternative way to use spacetime equations of motion to explore the finite entropy of the system, in a universal way insensitive to the details of such corrections. We can explore the resolution of the singularity by studying the system at a finite temperature, and determine its properties as a function of $N$ in the limit of vanishing temperature. We will find that the entropy goes as $N^{\frac12}$, in perfect agreement with the microstate counting of the previous section. This provides further evidence that the would-be stretched horizon of a stack of D0 branes should have an area that scales as $N^{\frac 12}$.

We take the near-extremal (finite temperature) solution for D0-branes (see e.g. \cite{Itzhaki:1998dd}), with a horizon $r_0$.   This comes down to introducing factors of 
\beqa f_0= \left(1- \frac{r_0^7}{r^7}\right)\eeqa in the metric \eqref{eq:D0einstein}, where $r_0$ is tied to the energy density $\epsilon$ of the branes above extremality:
\begin{equation} 
    \begin{gathered}
    r_0^{7}\sim  \alpha'^4 g_s ^2 \;\epsilon\,.
    \end{gathered}
\end{equation}
In the limit where $\epsilon \to 0 $, we recover the extremal (zero temperature) stack of D0 branes.

In the spirit of the gauge/gravity correspondence for general D-branes in \cite{Itzhaki:1998dd}, we will carry out the computation of the entropy in the near-horizon limit
\beqa 
r\to 0 \quad , \quad \alpha'\to 0\quad ,\quad U=r/\alpha'={\rm fixed}\quad ,\quad 
g_{QM}^2 = (2 \pi)^{-2} g_s {\alpha'}^{-3/2}= \text{fixed} \, .\quad\quad
\eeqa 
Here $g_{QM}$ is the coupling of the D0-brane worldvolume quantum mechanics. It is useful to define the dimensionless coupling at some energy scale $U$
\begin{equation}\label{geffdef}
    g_{\rm eff}^2 \sim g_{QM}^2 N U ^{-3}\sim \lambda U^{-3}\,,
\end{equation}
where $\lambda= g_{QM} ^2N$ is the (dimensionful) 't Hooft coupling.

The solution in the limit becomes
\beqa 
 &&  \frac{ds^2}{\alpha '}= -\frac{U^{7/2}}{\sqrt{d_0 \lambda}} f_0 \,dt^2 + \frac{\sqrt{d_0 \lambda}}{U^{7/2}} \left( \frac{dU^2}{f_0} +  U^{2} d\Omega_8^2 \right) \, , \nonumber \\
  &&  e^\phi = \frac{( 2\pi)^2}{d_0} \frac{1}{N} \left(\frac{\lambda d_0 }{U ^3}\right)^{7/4} \sim \frac{g_{eff}^{7/2}}{N}\,,
\label{eq:deformedD0}
\eeqa 
where 
\begin{equation} \label{U0-definition}
    \begin{gathered}
    d_0 = 240 \pi^5\, , \quad 
        U_0^{7}= a_0\; g_{QM} ^4 \;\epsilon, \;\;\;\;\; a_0 =\frac{4480\, \pi^7}{3}\,.
    \end{gathered}
\end{equation}

The horizon $U_0$ can be expressed in terms of the temperature as: 
\begin{equation}
    \label{eq:temperature} T^{-1}= \frac{4}{7} \pi \sqrt{\lambda d_0} U_0 ^{-5/2}\,.
\end{equation}
We now move on to the Einstein frame to compute the entropy. The Einstein frame spatial metric reads \cite{Anous:2019rqb}:
\begin{equation}
    ds^2_E = C U^{-7/8} \left( \frac{dU^2}{f_0} + U^2 d\Omega_8^2\right) \;\; {\rm with }\;\;
    C= \frac{l_s^2 N^{1/2} d_0 ^{1/8}}{2 \pi \lambda ^{3/8}}\, .
\end{equation}
It is convenient to introduce a new radial variable $R$, with dimension of length, defined as $U^{9/8}= (R^2/C)$. The metric becomes: 
\begin{equation}
    ds^2_E = \frac{256}{81} \frac{dR^2}{1-(\frac{R_H}{R})^{112/9}} + R^2 d\Omega_8^2\, ,
\end{equation}
where $R_H$ has the parametric dependence
\beqa 
  R_H \sim \left(\frac{T}{\lambda^{1/3}}\right)^{9/40} N ^{1/4} l_s\,.
\eeqa 
So the entropy of the finite temperature black hole in terms of the dimensionless temperature $T/ \lambda^{1/3}$ is given by
\begin{equation} \label{eq:finite-temperature-entropy}
    S \sim N^2 \times (T/ \lambda^{1/3})^{9/5} \, .
\end{equation}
Restoring the powers of $N$ hidden in $T$ and $\lambda$, we get the scaling behaviour
\begin{equation} \label{entropy-non-extremal}
    S\sim  (U_0^{5/2}/g_{QM}^{5/3} )^{9/5} \, N^{1/2} \, .
\end{equation}
In the limit $U_0\to 0$, one recovers the familiar zero entropy result. However, the point that the above computation shows is that entropy of the system, when cloaked with a fixed horizon size, scales as $N^{1/2}$, in agreement with our discussion in section \ref{sec:microscopic}.

\section{Species Scale Horizon from Emergent Higher Derivative Terms}
\label{sec:final-horizon-d0}

It is useful to have a more direct intuition, at the level of the geometry, about the appearance of the species scale horizon for D0-branes uncovered in the previous section. In analogy with the 4d small black holes in section \ref{sec:tower-4d}, we expect this horizon to show up once higher derivative corrections are included in the effective theory, which become relevant precisely at the species scale. Moreover, such new terms are expected to arise from integrating out the towers of light particles associated to the infinite distance limit of the formerly small black hole, in the spirit of Emergence.

In this section we work out in detail this picture for the D0-brane system. We first show that generic 10d $R^4$ corrections lead to a finite size horizon. We then consider the specific $R^4$ couplings arising from integrating out a tower of D0-brane states (i.e. 11d KK gravitons), which constitute the SDC tower associated to the infinite distance limit probed by the small black hole solution (decompactification to 11d M-theory). Interestingly we find that these familiar $R^4$ terms do not suffice to generate the finite size horizon for the system (let alone a species scale one). Turning things around, this suggests that the appearance of the species scale horizon demanded by the swampland considerations requires a crucial role of new physics beyond the usually considered $R^4$ terms. In fact, we include a specific set of supersymmetric 8-derivative terms involving the RR field strength 2-form, tractable via a lift to 11d Mtheory, and show that they do  produce the species scale horizon, thus confirming the swampland expectations. 

\subsection{Entropy function analysis of general $R^4$ terms}
\label{sec:entropy-functional-d0}

It is well established that the existence and properties of stretched horizons for 4d small black holes in theories with higher-curvature corrections can be studied using the entropy function formalism \cite{Sen:2005wa}. The idea is to use the ansatz for a putative near horizon AdS$_2$ solution to evaluate an entropy function, and to extremize it with respect to its parameters. In this section we apply a similar logic to the solution of D0-branes in 10d type IIA, and argue that the introduction of higher curvature corrections leads to a stretched horizon for the system.

Before entering the discussion, let us make a general comment. In this section we only include general terms involving only the curvature, but no other fields. This is for tractability, since curvature terms lead to closed expressions which can be analyzed in a model-independent way, and suffice to illustrate the fairly generic appearance of stretched horizons. The discussion of other corrections can be carried out for instance if a specific set of such correction is fixed, as we do in section \ref{sec:mtheory-story} for a set of 8-derivative corrections including curvature and the RR 2-form field strength.

\subsubsection{The D0-brane solution at two-derivative level}
\label{sec:d0-two-derivative}

We start by recovering the key features of the two-derivative solution of the D0-brane system from the entropy function formalism \cite{Sen:2005wa}, as warmup for coming sections. As explained above, we introduce an AdS$_2\times\IS^8$ ansatz:
\begin{equation} \label{d0-ansatz} 
\begin{split} 
      ds^{2} &= v_1 \left( -r^2 dt^2 + \frac{1}{r^2} dr^2 \right) + v_2 d\Omega_{8}^{2} \, , \\
	  F_{rt} &= e \, , \quad \tilde F_{\theta \phi_1 \cdots \phi_7} = 0 \, , \\
	  e^{\phi} &= g_s \, ,
\end{split}
\end{equation}
where $v_1$ and $v_2$ control the AdS and sphere length scales. For the 2-form field strength this introduces electric but not magnetic charge at the horizon. Finally, the last equation sets the value of the string coupling at the horizon.

The entropy function is given (modulo some irrelevant overall factor) by evaluating the action at the near horizon ansatz (\ref{d0-ansatz}) and applying a Legendre transform with respect to $e$, namely
\begin{equation} \label{entropy-functional}
  \mathcal E (N,g_s,v,\beta,e) = e N - \int_{S^8} d\Omega_8 \sqrt{-g} \, \left.\mathcal L\right|_{h} \, .
\end{equation}
To evaluate this quantity for the ansatz \eqref{d0-ansatz} we compute
\begin{align} 
	\sqrt{-g} &= v_1 v_{2}^{4} \, \sqrt{g_{\Omega_8}} = \frac{v^{5}}{\beta} \, \sqrt{g_{\Omega_8}} \, , \label{ansatz-2sqrt} \\
	R &= - \frac{2}{v_1} + \frac{56}{v_2} = \frac{56 - 2\beta}{v} \, ,\label{ansatz-2R} \\
	|F_2|^{2} &= \frac{1}{2!} F_{\mu\nu} F^{\mu \nu} = -\frac{1}{v_{1}^{2}} e^2 = - \frac{\beta^{2}}{v^2} e^2 \,  . \label{ansatz-2F}
\end{align} 
With hindsight, we have introduced new variables $v=v_2$ and $\beta= v_2/v_1$, to describe the overall length scale of AdS$_2\times\IS^8$ and the relative scale separation.
The result is 
\begin{equation} \label{Wald-1}
   \mathcal E (N,g_s,v,\beta,e) = e N-\frac{8 \pi ^4 v^3 \left( \beta ^2 e^2 g_{s}^{3/2}-4 (\beta -28) v\right)}{105 \beta } \, .
\end{equation}

The goal is to extremize this function with respect to $g_s$, $v$, $\beta$ and $e$, expressing the result in terms of the D0-brane charge $N$. We start with the extremization with respect to $e$, which gives
\begin{equation}
  \frac{\partial \mathcal E}{\partial e} = 0 \quad \rightarrow \quad e = \frac{105 N}{16 \pi ^4 \beta v^3 g_{s}^{3/2}} \, .
  \label{extremization-e}
\end{equation}
Substituting this result, the entropy function reads 
\begin{equation}
  \mathcal E (N,g_s,v,\beta) = \frac{105 N^2}{32 \pi ^4 \beta v^3 g_{s}^{3/2}} +\frac{32 \pi^4 (\beta -28) v^4}{105 \beta } .
\end{equation}
We now extremize with respect to $v$ and $\beta$, and get
\begin{equation} \label{sol-v-beta}
  \frac{\partial \mathcal E}{\partial v} =  \frac{\partial \mathcal E}{\partial \beta} = 0 \quad \rightarrow \quad v^7 = \frac{1575 N^2}{4096 \pi ^8 g_{s}^{3/2}} \, , \ \beta = 49 \, .
\end{equation}
Incidentally, the approximately order 1 value of $\beta$ (taking volume factors into account) is natural from the absence of scale separation in AdS vacua \cite{Lust:2019zwm}. The resulting entropy function now reads
\begin{equation} \label{Wald-2}
  \mathcal E (N,g_s) \sim \frac{N^{8/7}}{g_{s}^{6/7}}\, .
\end{equation}

We see that there is no well-defined extremum for the parameter $g_s$, it lies at $g_s\to\infty$. This simply reflects the fact that the D0-brane solution drives the dilaton to infinity in moduli space at its core, as observed above from the 10d solution. Indeed, we can recover other features of the 10d solution from the entropy function, by studying the scaling of different quantities with $g_s$. For instance, from \eqref{sol-v-beta} and \eqref{ansatz-2R}, we have that in the $g_s \to \infty$ limit 
\begin{equation}
  v_2 = v \sim \frac{1}{g_{s}^{3/14}} \to 0 \,, \quad R \sim g_{s}^{3/14} \to \infty\,  .
\end{equation}
Namely,  we recover that the horizon volume goes to zero, and the scalar curvature blows up with exactly the same scalings as would be obtained directly from the D0 solution at its core. Note that the $g_s$ we consider in this section is to be compared to the dilaton near the core of the D0 stack in the solution \eqref{eq:D0einstein}. Indeed, $g_s$ is the value of the string coupling at the horizon of the would-be black hole, whilst $g_s^\infty$ in \eqref{eq:D0einstein} is the string coupling infinitely far away from the stack of branes. 

We therefore recover that the process of extremization of the Wald entropy function of the near-horizon geometry of a black hole electrically charged under the RR 2-form in 10d describes the near-core behaviour of a stack of D0 branes. The fact that the Wald entropy is only extremized for a value of the string coupling at the core that is blowing up, with a horizon that tends to zero size is exactly the behaviour expected from small black holes, as in Section \ref{sec:stretched-4d}. We conclude from this that the Wald entropy formalism constitutes a reliable description of D0-brane stacks as small black holes. In the next section, we will introduce higher derivative corrections to the 10d supergravity which will affect the Wald entropy in such a way that the same exact minimization procedure will lead to a finite sized black hole instead of a small one. 

\subsubsection{The D0-brane solution and higher derivative terms}
\label{sec:d0-four-derivative}

In analogy with the 4d small black hole case, we may expect that the D0-brane solutions can develop a stretched horizon upon the inclusion of suitable higher derivative corrections. In 10d type IIA the first corrections allowed by supersymmetry arise at 8-derivative level. There is a vast literature devoted to the computation of these corrections at tree and one-loop level (see e.g. \cite{Green:1981yb,Gross:1986iv,Grisaru:1986vi,Grisaru:1986dk,Grisaru:1986kw,Green:1997as,Grimm:2017okk}). However, for the present purposes it will be more efficient to assume the presence of general $R^4$ corrections (the example of supersymmetric $R^4$ corrections will be discussed explicitly in section \ref{sec:r4terms}). Also, as explained earlier, we focus on terms involving only curvatures, which allows for a tractable model-independent analysis. The inclusion of terms involving other fields is discussed in section \ref{sec:mtheory-story} for a specific set of supersymmetric 8-derivative terms including the RR 2-form field strength.

We now describe the entropy function computation in the presence of general $R^4$ corrections. The corrections can be written as a linear combination of all possible contractions between eight inverse metrics and four Riemann tensors with all the indices down. Inspecting the ansatz in \eqref{d0-ansatz}, we see that each inverse metric gives either a $1/v_1$ or a $1/v_2$ contribution, while each Riemann gives either a $v_1$ or a $v_2$. In terms of $v$ and $\beta$, this means that any of these contributions will take the form $\beta^{a}\, v^{-4}$, with $a$ ranging from $-4$ to $8$. For instance, the $a=-4$ one corresponds to the case in which all the Riemanns give $v_1$ and all the metrics give $1/v_2$, while the $a=8$ one corresponds to all the Riemanns giving $v_2$ and all the metrics giving $1/v_1$. 

Following this reasoning, we can write the $R^4$ term evaluated at the horizon as
\begin{equation} \label{R4-nearhorizon}
  \left. \mathcal L_{R^4} \right|_{h} = - g_{s}^{1/2} \ \frac{p_{12}(\beta)}{\beta^4} \, \frac{1}{v^4} \, ,
\end{equation}
where $p_{12}$ denotes a degree 12 polynomial, and the minus sign is there for future convenience. Note that we already make explicit the dependence on the string coupling, which is fixed to that of the $R^4$ terms arising at first loop in perturbative Type IIA string theory. Motivated by having $g_s \to \infty$ at the core of the D0 solution at the two-derivative level, we are neglecting the tree level piece. One could worry that even higher loop contributions would dominate as $g_s \to \infty$, but they actually vanish. This has been considered previously in the literature from the Type IIA perspective \cite{Green:1997di,Green:1997as,Russo:1997mk} and, as it will become clearer in section \ref{sec:mtheory-story}, it is required for a consistent uplift to M-theory in this limit. 

Including the extra term in \eqref{R4-nearhorizon}, the entropy function \eqref{entropy-functional} reads
\begin{equation}
  \mathcal E (N,g_s,v,\beta,e) = e N -\frac{8}{105} \pi ^4 \beta  e^2 g_{s}^{3/2} v^3 +\frac{16 \pi ^4 v \left(g_{s}^{1/2} p_{12}(\beta) +2 (\beta -28) \beta ^4 v^3\right)}{105 \beta ^5} \, .
\end{equation}
The extremization with respect to $e$ is insensitive to the new $R^4$ term, so the result is again \eqref{extremization-e}.
Plugging this result, the entropy function reads
\begin{equation}
  \mathcal E (N,g_s,v,\beta) = \frac{105 N^2}{32 \pi ^4 \beta  g_{s}^{3/2} v^3}+\frac{16 \pi ^4 \left(g_{s}^{1/2} p_{12}(\beta) v+2 (\beta -28) \beta ^4 v^4\right)}{105 \beta ^5} \, .
\end{equation}

In order to display the effect of the higher derivative term in the dilaton, let us now consider the extremization with respect to $g_s$. We obtain
\begin{equation} \label{eq:ext-gs}
  \frac{\partial \mathcal E}{\partial g_s} = 0 \quad \rightarrow \quad g_s = \frac{105}{16 \pi ^4 } \sqrt{\frac{3}{2}}\frac{\beta ^2 N}{\sqrt{p_{12}(\beta)}v^2}\, .
\end{equation}
We see that the new contributions generically stabilize the dilaton. We note that the solution only exist if $p_{12}(\beta_0) > 0$, where $\beta_0$ is the yet to be determined value for $\beta$ extremizing the entropy function. The latter is determined by the extremization with respect to $v$, namely
\begin{equation} \label{sol-beta}
  \frac{\partial \mathcal E}{\partial v} = \frac{128 \pi ^4 (\beta -28) v^3}{105 \beta } =0 \quad \rightarrow \quad \beta = 28 \, .
\end{equation}
Note that we again obtain a value compatible with the absence of scale separation, suggesting that this property is fairly robust against inclusion of higher derivative corrections.

Finally, the extremization with respect to $\beta$ gives
\begin{equation} \label{sol-v}
  \left. \frac{\partial \mathcal E}{\partial \beta} \right|_{\beta=28}= 0 \quad \rightarrow \quad v^4 = \frac{ 4\, p_{12}(28)-21 \,p_{12}^{\prime}(28) }{175616 \cdot 6^{1/4} \, \pi^2 \, p_{12}(28)^{1/4}} \, \sqrt{\frac{5 N}{7}} \, ,
\end{equation}
where $p_{12}^{\prime}(\beta)$ is the derivative of the polynomial with respect to $\beta$. Hence, the existence of the solution requires 
\beqa 
p_{12}(28) > 0\quad , \quad 4\, p_{12}(28)>21 \,p_{12}^{\prime}(28)\, .
\label{the-conditions}
\eeqa
We will see the explicit evaluation of such polynomials for a particular $R^4$ correction in section \ref{sec:r4terms}.

In conclusion, under fairly general circumstances, the presence of higher derivative corrections suggests the existence of a stretched horizon for 10d D0-brane solutions. From the above expressions, the scaling of its properties with $N$ is 
\beqa
	S \,\sim \,\sqrt{N} \quad , \quad 
	v^4 \,\sim \, \sqrt{N} \quad , \quad
	\beta \,= \,28 \quad , \quad
	g_s \,\sim\, N^{3/4} \quad , \quad
	e \,\sim\, N^{-1/2} \, .
 \label{near-horizon-scalings}
\eeqa
The entropy of the system scales as $\sqrt{N}$, in agreement with the microstate counting in section \ref{sec:microscopic}. Notice that this, as well as all the scalings above, rely heavily in the string coupling dependence made explicit in \eqref{R4-nearhorizon} and corresponding to the one-loop contribution in perturbative Type IIA string theory. The string coupling at the horizon blows up as $N\to \infty$, which is consistent with neglecting the tree level contribution in this regime. In the next section we will shed some further light on these scalings, since they lead to the appearance of the species scale. 

\subsubsection{The Species Scale and Cosmic Censorship}
$\,$
\vspace{-15pt}

{\bf The Species Scale}

Here we shortly note that the above horizon scale indeed corresponds to the species scale of the SDC tower associated to the infinite distance limit probed by the solution. This infinite distance limit corresponds to a decompactification to 11d M-theory (consistently with the above mentioned fact that the horizon value of $g_s$ gets strong in the large $N$ regime). The extra dimension is reconstructed by the SDC tower of D0 branes, whose species scale corresponds to the 11d Planck scale. Restoring the 10d Planck scale, we have
\begin{equation}
    \Lambda_{s} \sim M_{11} \sim g_{s}^{-1/12} M_{10} \sim N^{-1/16} M_{10}\, ,
\end{equation}
where we have used the scaling of $g_s$ with $N$ in equation \eqref{near-horizon-scalings}. It is indeed easy to show that this coincides with the inverse radius of the stretched horizon. Using the scaling of $v$ with $N$ in \eqref{near-horizon-scalings}, we have
\begin{equation}
    r_{h}^{-1} \sim v^{-1/2} M_{10} \sim N^{-1/16} M_{10} \, .
\end{equation}
Let us note that this agreement is easily understood by the fact that the stretched horizon arises from the competition between the classical and the $R^4$ terms in the effective action. The scale at which these two can compete indeed corresponds to one of the notions of the species scale \cite{vandeHeisteeg:2022btw}. The fact that it matches the species scale associated to the tower of D0s can be directly checked in the effective action and, as we explain in section \ref{sec:tower-10d}, can be understood from the emergence of higher derivative terms coming from integrating out the tower of D0s. Here we see that these two notions of species scale also coincide with the one of the smallest BH describable within the EFT, in this case the stretched horizon of the D0 black hole.

\bigskip

{\bf The D0-brane bound state size scale}

We have emphasized the appearance of a finite size for the system from the spacetime viewpoint, via the appearance of a stretched horizon. However, as any quantum system, the D0-brane system has a characteristic size, in particular that associated to the size of the D0-brane bound states. This has been studied in particular in the context of M(atrix) theory. In this context, the outcome of \cite{Banks:1996vh,Polchinski:1999br,Susskind:1998vk,Nekrasov:1999cg}, is an estimate of the size of a bound state of $N$ D0-branes scaling as
\beqa
R\gtrsim N^{1/3}M_{11}^{-1} \, . 
\label{bound-state-size}
\eeqa 
This grows with $N$ in a way seemingly stronger\footnote{Note that the stronger growth also holds even if we restrict to the dominant bound state of $k\sim N^{1/2}$ D0-branes, as in section \ref{sec:microscopic}, which would yield $R\sim N^{1/6}$.} than the stretched horizon size $r_h \sim M_{11}^{-1}$ discussed above. However, an important observation is that (\ref{bound-state-size}) is derived in the limit of $g_s\to 0$ with $M_{11}$ fixed, with $g_s$ the coupling in flat space. Hence, for a proper comparison, in the evaluation of the horizon size $r_h \sim N^{1/16} M_{10}^{-1}\sim g_{s}^{-1/12} N^{1/16} M_{11}^{-1}$, one should not use the horizon value $g_s\sim N^{3/4}$ in matching $M_{10}$, but rather consider $g_s$ as a free asymptotic parameter. Thus, in the $g_s\to 0$ and $M_{11}$ fixed limit, the horizon size is parametrically larger than the size of the bound state of gravitons, which are effectively cloaked behind the horizon.

\bigskip

{\bf A cosmic censorship interpretation}

The above considerations motivate a cosmic censorship interpretation of our result, which provides a different angle to other uses of cosmic censorship in the analysis of swampland constraints (see e.g. \cite{Crisford:2017gsb,Horowitz:2019eum}). The classical 2-derivative solution displays a singularity. In the full theory, the pathological implications of such singularities must be avoided, either by some desingularization \cite{Klebanov:2000hb}, addition of extra degrees of freedom as in orbifolds, or the appearance of a horizon cloaking it. We have shown that in our system the latter occurs via the use of higher derivative interactions, despite the fact that it does not happen in the low-energy two-derivative approximation. Interestingly, this happens in the realm of the effective field theory, on the verge of its validity. This is again tied up to the notion of the species scale as that at which the tower of higher derivative interactions becomes relevant.

\subsection{The D0-brane SDC tower and Higher Derivative Emergence}
\label{sec:tower-10d}

The infinite distance limit in moduli space explored by the core of the solution is the strong coupling limit of type IIA theory, which corresponds to the decompactification limit of 11d M-theory on $\IS^1$. The tower of states predicted by the swampland distance conjecture corresponds to D0-brane states\footnote{We expect readers not to confuse the D0-brane states in the SDC tower with the D0-branes sourcing the solution in the previous sections.}, namely the KK momentum states of 11d graviton multiplets. The computation is indeed simpler in this 11d picture, and is a higher-dimensional version of that encountered in section \ref{sec:1loop-4d}. In fact is a well-established story that this set of states can lead, at the 1-loop level, to the appearance of higher derivative corrections to the 10d effective action, which can be interpreted as tree- and one-loop level (in $g_s$) terms of type IIA theory \cite{Green:1997as}, as we now discuss.

The terms we are going to compute are the $t_8t_8R^4$ terms\footnote{As we will explain in section \ref{sec:r4terms}, the actual 10d type IIA $R^4$ terms contain a further piece with an $\epsilon_{10}\epsilon_{10}$ structure. This however does not contribute to 4-graviton scattering amplitudes, or conversely is not captured by our 4-graviton 1-loop amplitude (it would require a 5-point computation). This is analogous to the similar statements for the odd-odd structure in the $R^4$ computation from scattering amplitudes in string theory, see e.g. \cite{Liu:2013dna}.}, in principle of the kind included in the analysis in section \ref{sec:d0-four-derivative} (see also section \ref{sec:r4terms} for more details on their structure). They are protected by supersymmetry, so only the 1-loop diagram of BPS 11d graviton KK modes can contribute. These can be easily computed from a worldline formalism. Using the general expression (\ref{general-amp}), integrating over continuous momenta, and particularizing for $d=10$, $n=4$, we get
\begin{equation}
    {\cal A}_{10,4} = \frac{\sqrt{\pi}}{2 R} \tilde{K} \int_{0}^{\infty} \frac{d\tau}{\tau^{2}} \sum_{k} e^{- \pi \tau R^{-2} k^2} \, ,
    \label{before-poisson}
\end{equation}
where the prefactor $\tilde K$ contains all the kinematics, and is the linearized approximation to the $t_8 t_8 R^4$ term in the effective action. As in the 4d case, the above expression diverges in two ways: because of the integral and because of the infinite sum. We can perform a Poisson resummation, so that both are nicely combined and we get
\begin{equation}
    \frac{1}{\pi^{3/2}}{\cal A}_{10,4} = \tilde{K} \int_{0}^{\infty} d\hat{\tau} \, \hat{\tau}^{1/2} \sum_{l} e^{- \pi \hat{\tau} R^{2} l^2} \,  = C \tilde{K} + \frac{\zeta(3)}{\pi R^3} \tilde{K} \, ,
    \label{eq:amplitude-result}
\end{equation}
where $\hat \tau = \tau^{-1}$. This Poisson summation trades the KK momentum $k$ for the winding number $l$ of the worldline along the $\IS^1$. The only divergence is now in the $l=0$ piece, which has been isolated as the first terms in the second equality; $C$ is an  unknown coefficient regularizing the divergence.\footnote{In \cite{Green:1997as}, it was fixed by T-duality upon further $\IS^1 $ compactification. In our present context, there is no way of fixing it without this kind of extra UV information. It would be interesting if performing the sum over KK modes up to the species scale one could reproduce its precise value, along the lines of \cite{Marchesano:2022axe,Castellano:2022bvr,Blumenhagen:2023yws,Blumenhagen:2023tev,Blumenhagen:2023xmk}.}

As anticipated, the quantum corrections coming from the SDC tower lead to a higher derivative $R^4$ correction, in principle of the kind invoked to lead to a stretched horizon. There remains to perform a detailed evaluation of these terms and their effect in the entropy function argument, which we postpone to section \ref{sec:r4terms}.

An important remark is that the $R^4$ correction obtained in this way is ambiguous. The 4-graviton scattering is an on-shell quantity that is invariant under local field redefinitions $g_{ij} \to g_{ij} + \delta g_{ij}$, which is usually used to get rid of terms involving Ricci tensors (see e.g. \cite{Sinha:2006yy}). However, since the AdS$_2 \times \IS^8$ ansatz has a non-vanishing Ricci tensor, doing this is not innocuous for our purposes. We will thus stick to the full $t_8 t_8 R^4$ term, including Ricci terms, as it is the natural kinematic function that appears in this computation \`{a} la Emergence. That this tensor structure gives the full off-shell $R^4$ correction to the effective action was also suggested in \cite{Grisaru:1986vi}. As evidence for this, it was found that the terms containing no more than one Ricci tensor, which were reliably computed from the vanishing of the worldsheet beta function, precisely match the ones appearing in the $t_8 t_8 R^4$ structure. 

\bigskip

{\bf Higher Derivative Emergence}

When translated to the Type IIA frame, the first and second terms in \eqref{eq:amplitude-result} precisely reproduce the $t_8t_8 R^4$ tensor structure of the 1-loop and tree-level string perturbation theory contributions, respectively. Therefore, this establishes that these can be interpreted as coming from integrating out the full tower of D0-particles, as well as the fast-movers in the massless 10d sector. Let us remark that this result is very much in the spirit of the Emergence Proposal, albeit for higher derivative terms rather than just the kinetic terms. 

Note also that the above computation integrates out the full tower of states, i.e. not just up to the species scale as in \cite{Marchesano:2022axe,Castellano:2022bvr,Blumenhagen:2023yws,Blumenhagen:2023tev,Blumenhagen:2023xmk}. In fact, interpreting the latter recipe as a regularization procedure to be applied when integrating out infinite number of states, one could hope to recover the right value  for $C$; we leave this as an interesting open question. On the other hand, notice that integrating the full tower is in the spirit of the recent proposals in \cite{Blumenhagen:2023tev,Blumenhagen:2023xmk}. It would be interesting to explore these connections further.

\subsection{Supersymmetric 10d $R^4$ terms are not enough}
\label{sec:r4terms}

In this section, we particularize the analysis in section \ref{sec:d0-four-derivative} to the specific $R^4$ terms appearing in the Type IIA effective action (for a concrete choice fixing all the already mentioned ambiguities concerning these terms). By plugging the AdS$_2 \times$S$^8$ ansatz, we will compute the polynomial appearing in equation \eqref{R4-nearhorizon} and check the conditions in equation \eqref{the-conditions}.

We will show that, perhaps surprisingly, these supersymmetric $R^4$ terms do not satisfy the constraints in section \ref{sec:d0-four-derivative} required to generate a finite size horizon. Note that this does not invalidate the analysis of the entropy function, but rather shows that the appearance of the horizon from pure higher curvature terms demands the use of other possibly non-supersymmetric terms. Alternatively, one can maintain the problem in the realm of supersymmetric corrections, and include higher derivative couplings involving the RR 2-form field strength; these go beyond the analysis in section \ref{sec:d0-four-derivative}, and will be discussed in section \ref{sec:mtheory-story}.

Hence, we focus on the supersymmetric $R^4$ corrections in 10d type IIA theory.
Following the notation in \cite{Grimm:2017okk}, at the eight derivatives level, the type IIA action gets supplemented by additional terms quartic in the Riemann tensor at both tree level and first loop in the string coupling $g_s$:
\begin{equation}
    S_{\text{IIA}} = S_{\text{IIA}}^{class} + \frac{1}{3 \cdot 2^{11}}( S^{tree}+ S^{loop})\, ,
\end{equation}
where $S^{tree}$ and $S^{loop}$ both arise at level $\alpha'^3$.
In fact, it has been conjectured that these are the only two contributions to this type of term in string perturbation theory \cite{Green:1997di,Green:1997as,Russo:1997mk}. They can be written as follows in the Einstein frame and in 10d Planck units:
\begin{align}
    S^{tree}&= \frac{\zeta(3)}{2} \int d^{10}x \sqrt{-g} \, e^{- \frac{3}{2} \phi} \left( t_8 t_8 R^4 + \frac{1}{8} \epsilon_{10} \epsilon_{10} R^4 \right) \, ,\\
    S^{loop}&= \frac{\pi^2}{6} \int d^{10}x \sqrt{-g} \, e^{ \frac{1}{2} \phi} \left( t_8 t_8 R^4 - \frac{1}{8} \epsilon_{10} \epsilon_{10} R^4 \right) \, ,
\end{align}
where the form of the $t_8 t_8 R^4$ and $\epsilon_{10} \epsilon_{10} R^4$ structures can be found in Appendix \ref{appendix}. Let us only consider the 1-loop term, which is the leading one as $g_s \to \infty$.

Evaluating $t_8 t_8  R^4 $ and $\frac{1}{8} \epsilon_{10} \epsilon_{10} R^4$ on the $AdS_2 \times S^8$ near-horizon geometry \eqref{d0-ansatz}, one obtains: 
\begin{align} \label{eq:R4-typeIIA}
   t_8 t_8  R^4  & = \frac{192 \left(3 \beta^4 + 56 \beta^2 + 2520 \right)}{v^4} \, ,\\
   \frac{1}{8} \epsilon_{10} \epsilon_{10} R^4 & = \frac{161280 \left(4 \beta - 1 \right)}{v^4} \, .
\end{align}
Therefore, the 1-loop term  can be put in the form \eqref{R4-nearhorizon} if
\begin{equation}
    p_{12}(\beta) \sim - \beta^4 \left( 3 \beta^4 + 56 \beta^2 - 3360 \beta + 3360 \right) \, ,
\end{equation}
where we are ignoring an irrelevant positive-defined prefactor. Now we can check the conditions in \eqref{the-conditions} for the existence of an extremum of the entropy function. The second one is satisfied, while the first one is not. This implies that the familiar $R^4$ terms used so far do not suffice to generate the stretched horizon for the D0 solution.

This does not invalidate the main claim that the system should develop a species scale horizon. As explained earlier, we expect other higher derivative couplings to contribute non-trivially, and the horizon may very well develop when they are ultimately included. Indeed, we will show in the next section that a simple modification of the above computation promoted to M-theory allows to include couplings involving the RR 2-form $F_2$, and lead to a non-trivial species scale stretched horizon for the D0-brane system.

\subsection{Including $F_2$: Species scale D0-brane horizon from M-theory couplings}
\label{sec:mtheory-story}

In the previous section we showed that 10d supersymmetric $R^4$ terms do not generate a finite size horizon. Hence, the requirement that there should arise species scale horizon, as demanded by swampland considerations, implies that one should look further. One possibility is to consider other possibly non-supersymmetric pure higher curvature terms. A more attractive avenue is to stay in the realm of supersymmetric terms but include higher derivative couplings involving the RR 2-form field strength $F_2$. These go beyond the analysis in section \ref{sec:d0-four-derivative}, but, once a specific set of couplings is fixed, are amenable to a direct study again via the entropy function. 

In this section we include such terms, completing the analysis at the supersymmetric eight derivatives level of the effective action. The most efficient way of doing this is by encoding these terms in the 11d M-theory effective action, since both the 10d metric and RR 1-form become part of the 11d metric. This is, by encoding the extremal near-horizon ansatz in the 11d metric and plugging it into the M-theory effective action, supplemented by $R^4$ terms, we can effectively recover the 10d Lagrangian evaluated at the horizon including both $R^4$ and terms involving $F_2$. Let us remark that we will keep the 10d point of view at all times to focus in the presence of a stretched horizon for the D0 solution. This is, here we think about the 11d effective action as just an efficient way of encoding the eight derivatives terms involving $F_2$ and evaluating them at the extremal near-horizon ansatz.

Our analysis is close to that in \cite{Sinha:2006yy}, with the difference that we emphasize the 10d perspective in the computation. Even though we find analogous results, an important difference in the analysis is that we keep the full $t_8 t_8 R^4$ structure, while only purely Riemannian terms were included in \cite{Sinha:2006yy}. As already discussed in section \ref{sec:tower-10d}, the Wald entropy analysis is sensitive to the terms including the Ricci tensor which, via Emergence, are naturally encoded in the $t_8 t_8 R^4$ structure. This is consistent with the results obtained in \cite{Grisaru:1986vi}, where the terms in $t_8 t_8 R^4$ including one Ricci tensor were unambiguously shown to appear in the Type IIA effective action. 

Following again the notation in \cite{Grimm:2017okk}, and setting the three-form to zero for our purposes, the M-theory effective action can be written at the eight derivatives level as
\begin{equation} \label{eq:M-th-R4}
    S_{M} = \frac{1}{2 \kappa_{11}^{2}} \int d^{11}x \sqrt{- \hat g} \left( \hat{R} + \frac{\left(4 \pi \kappa_{11}^{2}\right)^{2/3}}{(2 \pi)^4 \cdot 3^3 \cdot 2^{13}} \left( \hat{t}_{8} \hat{t}_{8} \hat{R}^4 - \frac{1}{24} \epsilon_{11} \epsilon_{11} \hat{R}^4 \right) \right) \, .
\end{equation}
To simplify the computations, we will take $\kappa_{11}=1$.\footnote{Even though this might seem to imply that the results will be in 11d Planck units, when mapping to the type IIA Einstein frame, the 11d Planck scale gets traded for the 10d one. Thus, the final results are actually in 10d Planck units.} The $\hat{t}_8$ and $\epsilon_{11}$ tensors are defined in complete analogy with the Type IIA case. In fact, after writing these terms as contractions of four Riemann tensors, the 11d $R^4$ term look exactly the same as the 10d one. More details about these structures can be found in Appendix \ref{appendix}.

As it is well-known, the circle compactification ansatz from 11 to 10 dimensions is:
\begin{equation} \label{eq:11d-to-10d}
    d \tilde{s}^2 = e^{-\frac{1}{6} \phi} ds^2 + e^{\frac{4}{3} \phi} \left( dz - C_\mu dx^\mu \right)^2 \, .
\end{equation}
Plugging this into the effective action recovers the 10d Einstein frame Type IIA effective action up to the eight-derivative level. Instead of doing this and then taking the extremal $AdS_2 \times S^8$ near-horizon ansatz, we perform the computation at the level of M-theory. That is, we first encode the $AdS_2 \times S^8$ ansatz into the 11d metric, and then we plug it into the 11d effective action. This allows us to read off the 10d Lagrangian evaluated at the horizon. For this last step, we need to take into account the relation between the 11d and 10d determinant of the metric appearing in the action. This gives the following relation:
\begin{equation} \label{eq:11d to TypeIIA}
    \mathcal L_{IIA}|_{h} = g_{s}^{-1/6} \mathcal L_{M}|_{11d \, ansatz} \, . 
\end{equation}
To encode the extremal $AdS_2 \times S^8$ near-horizon geometry in the 11d metric ansatz, we just plug equation \eqref{d0-ansatz} into \eqref{eq:11d-to-10d} to get
\begin{equation} \label{eq:11d-ansatz}
    d \tilde{s}^2 = - g_{s}^{-1/6} v_ 1 \left( r^2 dt^2 + \frac{1}{r^2} dr^2 \right) + g_ {s}^{-1/6} v_ 2 \, d\Omega_{8}^{2} + g_ {s}^{4/3} \left( dz \
- e \, r \, dt \right)^2 \, .
\end{equation}

Evaluated on this ansatz, the 11d Ricci scalar reads
\begin{equation}
    \hat R = g_{s}^{1/6} \, \frac{ \beta^2 e^2 g_{s}^{3/2} - 4 v (\beta - 28 )}{2 v^2} \, ,
\end{equation}
where we introduced new variables $v=v_2$ and $\beta= v_2/v_1$ as we did in section \ref{sec:d0-two-derivative}. When plugged into \eqref{eq:11d to TypeIIA}, this recovers the two-derivative piece of the Type IIA Lagrangian evaluated at the horizon (c.f. equation \eqref{Wald-1}).

Similarly, we evaluate the M-theory $R^4$ terms in this ansatz, getting
\begin{equation} \label{eq:new-eq}
\begin{split}
    \left( \hat{t}_{8} \hat{t}_{8} - \frac{1}{24} \epsilon_{11} \epsilon_{11} \right) \hat{R}^4 = \frac{3 g_{s}^{2/3}}{4 v^8} 
    \bigg( & 287 \beta^8 e^8 g_{s}^6-1392 \beta ^7 e^6 g_{s}^{9/2} v \\
    & + 32 \left(83 \beta ^2+308\right) \beta ^4 e^4 g_{s}^3 v^2 \\
    & - 768 \left(3 \beta ^3+28 \beta -280\right) \beta ^2 e^2 g_{s}^{3/2} v^3 \\
    & + 256 \left(3 \beta^4 + 56 \beta^2 - 3360\beta +3360\right) v^4 \bigg) \, .
\end{split}
\end{equation}
As a check, we can recover the Type IIA $R^4$ result from the previous section by setting $e=0$ and plugging \eqref{eq:new-eq} into \eqref{eq:11d to TypeIIA} (c.f. equation \eqref{eq:R4-typeIIA} with the $g_{s}^{1/2}$ factor appearing in \eqref{R4-nearhorizon}).

Putting these two results together in the 11d Lagrangian, using equation \eqref{eq:11d to TypeIIA} and plugging the result into the definition of the entropy function in \eqref{entropy-functional}, we finally obtain:
\begin{equation}
\begin{split}
    &\mathcal E(N,g_s,v,\beta,e) = \, e \, N \, - \, \frac{8 \pi ^4 v^3 \left(\beta ^2 e^2 g_{s}^{3/2} - 4 (\beta -28) v\right)}{105 \beta } \\
    &-\left(\frac{\pi}{2}\right)^{2/3} \frac{g_{s}^{1/2}}{2580480 \, v^3 \beta} \bigg( 287 \beta ^8 e^8 g_{s}^6-1392 \beta ^7 e^6 g_{s}^{9/2} v + 32 \left(83 \beta ^2+308\right) \beta ^4 e^4 g_{s}^3 v^2 \\
    & - 768 \left(3 \beta ^3+28 \beta -280\right) \beta ^2 e^2 g_{s}^{3/2} v^3 + 256 \left(3 \beta^4 + 56 \beta^2 - 3360\beta +3360\right) v^4 \bigg) \, .
\end{split}
\end{equation}
Here we have already performed the integral in equation \eqref{entropy-functional}, taking into account the determinant of the metric evaluated in the ansatz as given in \eqref{ansatz-2sqrt}.

The entropy function turns out to be rather involved, and we cannot extremize it analytically. Doing so numerically would require fixing $N$ to a set of different values, extremizing the entropy function for each of these and then fitting the dependence on $N$ of the various quantities ($g_s , v$, ...). Instead of doing so, let us perform a change of variables inspired by the scalings found in section \ref{sec:d0-four-derivative}. We define
\begin{equation} 
	g_s = \tilde g_{s} \, N^{3/4} \, , \quad
    v = \tilde v \, N^{1/16} \, , \quad
    \beta = \tilde \beta \, , \quad 
	e = \tilde e \, N^{-1/2} \, .    
\end{equation}
Introducing this into the entropy function, we get
\begin{equation}
    \begin{split}
    &\mathcal E(N,\tilde g_s,\tilde v,\tilde \beta,\tilde e) = \sqrt{N} \Bigg[ \tilde e \, - \, \frac{8 \pi ^4 \tilde v^3 \left(\tilde \beta ^2 \tilde e^2 \tilde g_{s}^{3/2} - 4 (\tilde \beta -28) \tilde v\right)}{105 \tilde \beta } \\
    &-\left(\frac{\pi}{2}\right)^{2/3} \frac{\tilde g_{s}^{1/2}}{2580480 \, \tilde v^3 \beta} \bigg( 287 \tilde \beta ^8 \tilde e^8 \tilde g_{s}^6-1392 \tilde \beta ^7 \tilde e^6 \tilde g_{s}^{9/2} \tilde v + 32 \left(83 \tilde \beta ^2+308\right) \tilde \beta ^4 \tilde e^4 \tilde g_{s}^3 \tilde v^2 \\
    & - 768 \left(3 \tilde \beta ^3+28 \tilde \beta -280\right) \tilde \beta ^2 \tilde e^2 \tilde g_{s}^{3/2} \tilde v^3 + 256 \left(3 \tilde \beta^4 + 56 \tilde \beta^2 - 3360 \tilde \beta +3360\right) \tilde v^4 \bigg) \Bigg] \, .
\end{split}
\end{equation}
This change of variables does the important job of factoring out all the $N$-dependence in the entropy function, which is very non-trivial since all the terms had to conspire to give the same overall $\sqrt{N}$ factor. Thanks to this, we can now focus on the terms in parenthesis and look for its $N$-independent extrema numerically. If an extremum exist, then we see that the scalings we found in section \ref{sec:d0-four-derivative} are automatically guaranteed. Furthermore, given that the overall factor in front of the entropy function is $\sqrt{N}$, the scaling of the entropy from the microscopic counting in section \ref{sec:microscopic} is also automatically guaranteed.

\begin{figure}[htb]
\begin{center}
	\subfigure[$g_s$ fixed]{\includegraphics[width=0.4\textwidth]{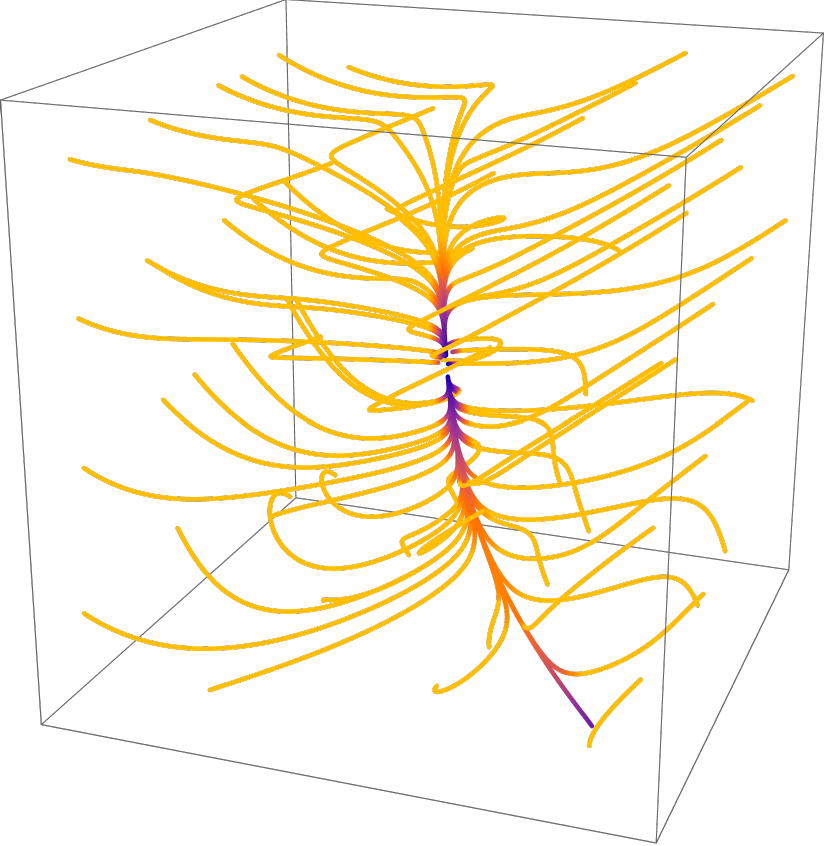}} \quad \quad
	\subfigure[$v$ fixed]{\includegraphics[width=0.4\textwidth]{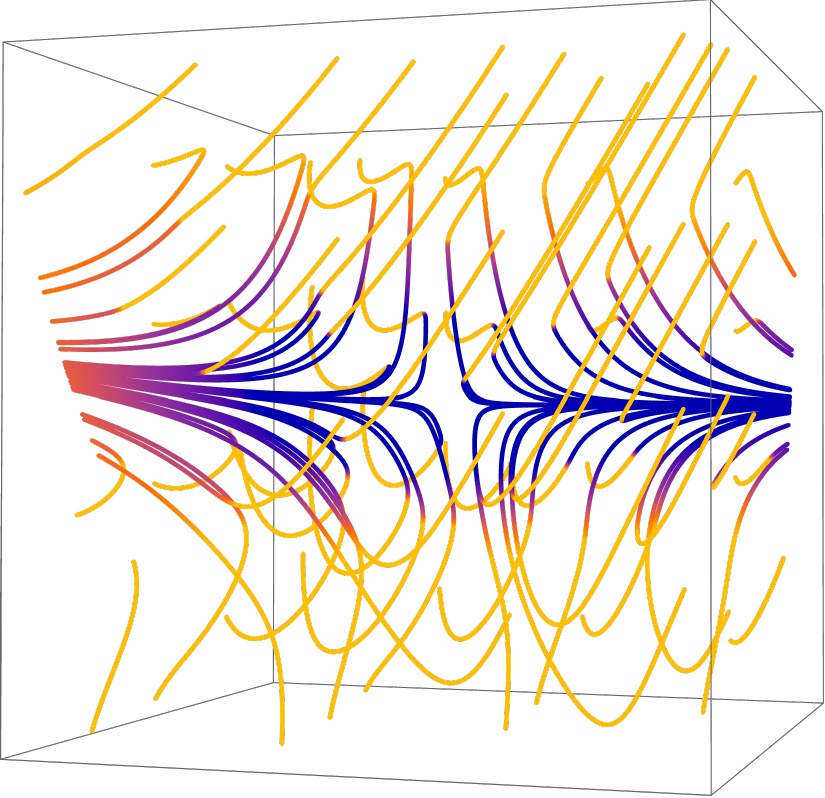}} \\
 	\subfigure[$\beta$ fixed]{\includegraphics[width=0.4\textwidth]{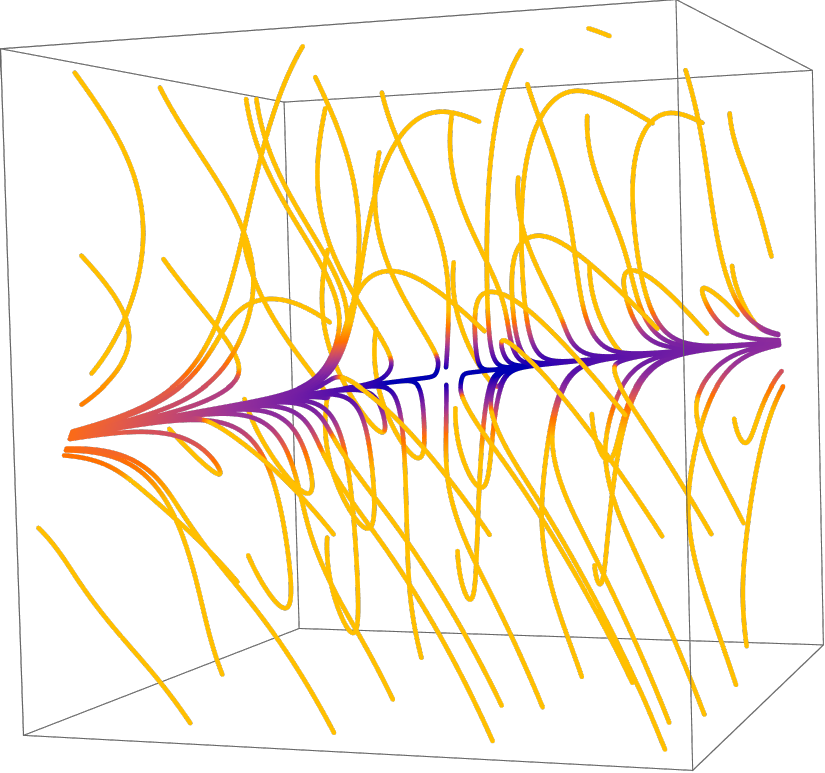}} \quad \quad
	\subfigure[$e$ fixed]{\includegraphics[width=0.4\textwidth]{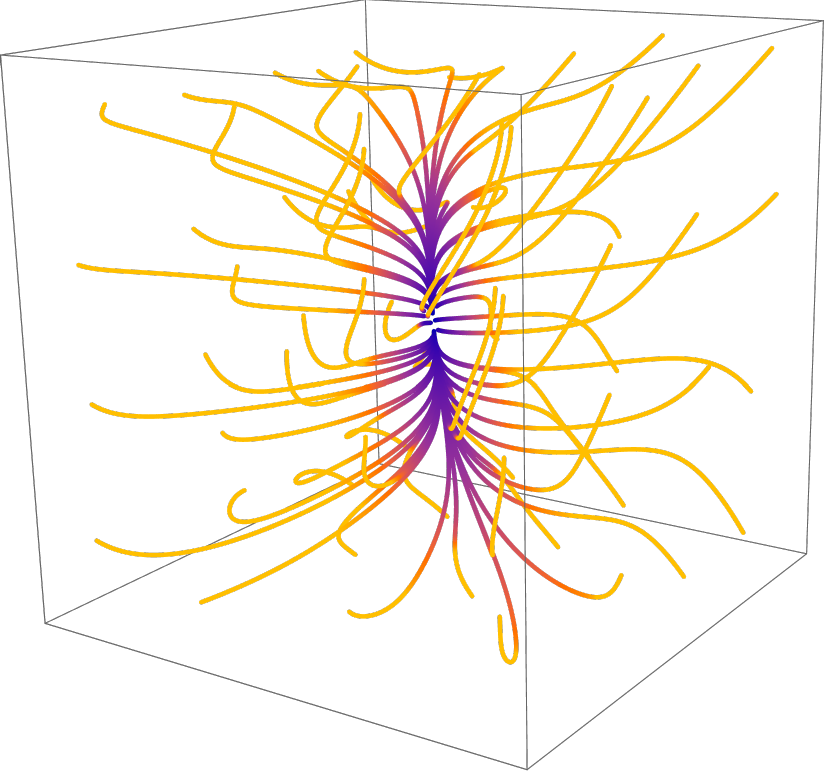}} 
	\caption{Gradient flows of the entropy function with one of the variables fixed. Lighter/yellow colors denotes bigger values for the gradient, while darker/blue colors denote smaller ones.}
	\label{fig:grad-flows}	
\end{center}
\end{figure}

Finally, we look for extrema of the terms in parenthesis in the previous equation or, equivalently, of $\mathcal E(1,\tilde g_s,\tilde v,\tilde \beta,\tilde e)$. This cannot be done analytically, so we perform a numerical search. The result is the following extremum:
\begin{equation}
    \tilde g_s \approx 2.90365 \, , \quad
    \tilde v \approx 0.159653 \, , \quad
    \tilde \beta \approx 14.0556 \, , \quad 
	\tilde e \approx 0.0479134 \, .    
\end{equation}
For these values, one can check that the derivatives of the entropy function are 14 orders of magnitude smaller than the value of function itself. This indicates that, to a very good approximation, this is an extremum of the entropy function. To make this extremum more apparent, we can fix one of the variables to its value at the extremum and display the three-dimensional gradient flow. This is shown in figure \ref{fig:grad-flows}, where we can see the presence of a non-trivial extremum. As it can be checked also by direct computation, we also see that the extremum is in fact a saddle point. This fact is not relevant for the Wald entropy formalism, that only requires the presence of an extremum. It would be interesting to understand the thermodynamical stability of this stretched horizon, somewhat along the lines of \cite{Sinha:2006yy}. However, this analysis presumably requires a systematic inclusion of even higher derivative terms to achieve control of order one factors. 

In conclusion, we found that the Type IIA action including \emph{all} terms up to the eight derivative level captures the presence of the species scale stretched horizon for the D0 solution. In addition, the entropy associated to the horizon nicely fits the microscopic counting and the near-horizon AdS$_2 \times$S$^8$ is not scale separated. This gives support to the idea that, even though a proper geometric treatment of a stretched horizon requires the inclusion of all the infinite amount of higher derivative operators, the classical level and the (complete) first non-trivial higher derivative level seem to suffice for capturing its presence and its scaling properties with the species scale.

As in the case of 10d $R^4$, the new set of higher derivative terms can be understood as arising from integrating out the SDC tower of D0-branes, in the spirit of Emergence. Indeed, since all these couplings are related by 11d Poincar\'e invariance, the computation of the 4-point amplitude is essentially unchanged. From the 10d perspective, the difference is that instead of computing 4-graviton scattering one should consider processes involving the RR 1-form field. Their worldline vertex operators are directly related \cite{Green:1999by}, hence the structure of the 1-loop integral is essentially unchanged, and  the changes only involve the kinematic tensor structure. This computation is directly not available in the literature, but it can be regarded as a 10d decomposition of the computation of the 11d $R^4$ correction from the 4-point 1-loop superparticle amplitude in \cite{Peeters:2005tb}.
This suffices to show the emergence from the SDC tower of the required higher derivative terms beyond the purely gravitational sector.
 
\section{Conclusions}
\label{sec:conclusions}

In this work we have clarified the links among the different avatars of the species scale in the swampland program: (i) as cutoff of the effective theory specially in the presence of the infinite SDC tower of light particles near infinite distance points in moduli space, (ii) the scale at which terms in the infinite series of higher derivative corrections in the gravitational sector start competing with the two-derivative level, (iii) as the size of the smallest black hole which admits a description within the EFT.

Our analysis uses small black holes, whose cores probe infinite distance limits in field space, at which we can characterize the SDC tower of states leading to the species scale in avatar (i). Upon integrating out the tower of states, the theory generates a series of higher-dimensional operators, naturally leading to avatar (ii) of the species scale. Finally, including these corrections in the black hole solution triggers the appearance of a stretched horizon of a size given by the species scale in avatar (iii).\footnote{It is worth noticing that we consider BPS small black holes, and that the quantum gravity effects we are taking into account come in the form of supersymmetry protected higher derivative corrections. It would be interesting to see how our results might extend to cases with less supersymmetry.}

We have provided quantitative examples of this phenomenon, both in a class of 4d small black holes, and in the fascinating case of 10d type IIA D0-branes. The latter case turned out to be highly non-trivial, as the appearance of the species scale horizon required understanding of 10d 8-derivative terms involving not only the curvature but also the RR 2-form field strength. The nice packaging of those terms into an 11d $R^4$ correction is a tantalizing hint that the species scale can serve as a powerful probe of the UV physics of M-theory.

The 10d D0-brane example displays an interesting feature: the objects making up the small black hole are precisely of the same kind as those in the SDC tower for the corresponding infinite distance limit. This feature arises in the microscopic explanation of the entropy for other black holes in string theory; but the remarkable fact about the D0-brane case is that the higher derivative terms coming from integrating out the tower of D0-branes are crucial to generate a geometric horizon for the D0-brane black hole itself!

This motivates us to entertain the following picture. According to the most radical interpretation of the Emergence Proposal \cite{Heidenreich:2017sim,Grimm:2018ohb,Heidenreich:2018kpg,Corvilain:2018lgw}, not only these higher derivative terms but the very dynamics of gravity would emerge from integrating out the towers of states. In this spirit, the emergence of species scale stretched horizons studied in this work would potentially apply to any black hole horizon. This suggests a possible general picture for the emergence of horizons in quantum gravity: In the UV, 
the natural quantum system to consider would be that formed by the states in the tower, with no dynamical gravity. When going down to the IR, the species in the tower are integrated out. It is reasonable to expect that this effective description cannot describe the previous physical system in full detail, but it should be able to codify its coarse-grained properties, i.e., its macrostate. How is this made possible? In the IR, gravity emerges and the system would be replaced by a black hole that precisely reproduces this macrostate. In this sense, the emergence of a horizon would be somehow required for the effective description to be able to keep the macroscopic information about the formerly considered physical system. 

It is also tantalizing to speculate that the entropy of the black hole is nothing but a reflection of the entanglement entropy of the species after their degrees of freedom are traced out. This goes in the spirit of ER=EPR \cite{Maldacena:2013xja}, the N-portrait picture \cite{Dvali:2011aa}, or of the literature on thermal M(atrix) theory black holes, see e.g. \cite{Banks:1997hz,Klebanov:1997kv,Horowitz:1997fr,Li:1997iz,Banks:1997tn,Susskind:1997dr,Li:1998ci} (also \cite{Smilga:2008bt,Wiseman:2013cda} for more recent developments). On a related note, it would be interesting to understand if and how this system made out of species could radiate and reproduce the evaporation of the black hole, perhaps along the lines of \cite{Cribiori:2023ffn}.

Back to more earthly matters, our work suggest several interesting questions for further research, for instance:

$\bullet$ We have focused on the interpretations of the species scale in asymptotic regions of moduli space, which are those naturally explored by small black holes. It would be interesting to extend our concept of
unification of the three notions of species scale and gain a similar understanding in the interior of moduli space, connecting with the proposal in \cite{vandeHeisteeg:2022btw,Cribiori:2022nke,vandeHeisteeg:2023ubh,Andriot:2023isc}, perhaps using large black hole probes as in \cite{Delgado:2022dkz}.

$\bullet$ Even in asymptotic regimes, there is a strong ongoing activity in understanding the precise treatment of the species scale in the process of integrating out the SDC tower of states, with different proposals \cite{Marchesano:2022axe,Castellano:2022bvr,Blumenhagen:2023yws,Blumenhagen:2023tev,Blumenhagen:2023xmk}. We hope the appearance of such computations in our setup can serve to clarify the proper physical procedures.

$\bullet$ The non-trivial constraints on higher derivative corrections for the appearance of stretched horizons are reminiscent of similar conditions derived from positivity constraints in the swampland program or in the S-matrix bootstrap, see e.g. \cite{Adams:2006sv,Cheung:2014ega,Bellazzini:2015cra,Hamada:2018dde}. It would be interesting to explore this possible connections further. 

$\bullet$ The AdS$_2\times \IS^8$ ansatz in the entropy function analysis suggests, via holography, the possible appearance of IR superconformal behaviour in the worldvolume theory on the D0-brane, once the equivalent to the higher derivative corrections are taken into account. It would be interesting to find further quantitative evidence for this proposal from the field theory side.

$\bullet$ It would be interesting to see whether some of our results extend to lower codimension cases. Higher-dimensional objects that probe infinite distance in field space and are singular in the EFT (such as the End-of-the-World branes of \cite{Buratti:2021yia,Buratti:2021fiv,Angius:2022aeq,Blumenhagen:2022mqw,Blumenhagen:2022bvh,Basile:2022ypo,Blumenhagen:2023abk}\footnote{For the related topic of solutions in theories with dynamical tadpoles, see \cite{Dudas:2000ff,Blumenhagen:2000dc,Dudas:2002dg,Dudas:2004nd} for early work and \cite{Raucci:2022jgw,Antonelli:2019nar,Mininno:2020sdb,Basile:2020xwi,Basile:2021mkd,Mourad:2022loy} for related recent developments.}) could, in the same way as the small black holes, be shown to develop a species scale sized horizon using a modified version of the entropy function formalism.

$\bullet$ In some instances, higher derivative corrections have been shown to create new singularities on the horizon of large black holes \cite{Horowitz:2023xyl}. This is to be put in contrast with our work, where curvature corrections instead smooth out singularities. It would be interesting to better understand the mechanisms that make it such that curvature corrections can seemingly both create and smooth out singularities in the EFT.

$\bullet$ There are interesting recent development regarding the possibility of log corrections to the species scale \cite{vandeHeisteeg:2023ubh,Cribiori:2023sch,Blumenhagen:2023yws}. In this respect, it is tantalizing that our microscopic considerations in section \ref{sec:microscopic} lead to such  corrections. It would be interesting to exploit our techniques to shed some light on this question.

We hope to come back to these questions in the near future.

\section*{Acknowledgments}
We thank Roberta Angius, Alberto Castellano, Lucia G\'omez Cordova, Jes\'us Huertas, Luis Ib\'a\~nez, Shota Komatsu, Wolfgang Lerche, Luca Melotti, Fernando Marchesano, Jake McNamara, Miguel Montero, Tom\'as Ort\'in and Irene Valenzuela for useful conversations. The work of M.D. and A.U. is supported through the grants CEX2020-001007-S and PID2021-123017NB-I00, funded by MCIN/AEI/10.13039/501100011033 and by ERDF A way of making Europe.

\appendix

\section{Tensor Structure of $R^4$ Corrections}
\label{appendix}

In this appendix we recall the form of the $t_8 t_8 R^4$ and $\epsilon_{10} \epsilon_{10} R^4$ terms arising in 10d type IIA theory (and closely related to those in 11d M-theory). We follow closely the notation in \cite{Grimm:2018ohb}.

Let us start with the $t_8 t_8 R^4$ structure, which denotes
\begin{equation} \label{eq:def-t8t8R4}
    t_8 t_8 R^4 = t_{8}^{A_1 \cdots A_8} t_{8}^{B_1 \cdots B_8} \, R_{A_1 A_2 B_1 B_2} \, \cdots \, R_{A_7 A_8 B_7 B_8} \, ,
\end{equation}
where the $t_8$ tensor can be written in terms of the metric as\footnote{Notice the change of prefactor with respect to equation (B.3) of \cite{Grimm:2018ohb}, so that, after expressing this structure as contractions of four Riemann tensors, the result matches that in \cite{Grisaru:1986vi}.}
\begin{equation}
\begin{split}
    t_{8}^{A_1 \cdots A_8} = \frac{1}{5} \bigg[ &-2 \left( 
        g^{A_1 A_3} g^{A_2 A_4} g^{A_5 A_7} g^{A_6 A_8} +
        g^{A_1 A_5} g^{A_2 A_6} g^{A_3 A_7} g^{A_4 A_8} +
        g^{A_1 A_7} g^{A_2 A_8} g^{A_3 A_5} g^{A_4 A_6} 
    \right) \\ &+ 8 \left( 
        g^{A_2 A_3} g^{A_4 A_5} g^{A_6 A_7} g^{A_8 A_1} +
        g^{A_2 A_5} g^{A_6 A_3} g^{A_4 A_7} g^{A_8 A_1} +
        g^{A_2 A_5} g^{A_6 A_7} g^{A_8 A_3} g^{A_4 A_1} 
    \right) \\
        &- (A_1 \leftrightarrow A_2) - (A_3 \leftrightarrow A_4) - (A_5 \leftrightarrow A_6) - (A_7 \leftrightarrow A_8)
    \bigg] \, .
\end{split}
\end{equation}
After contracting all the metrics in $t_8$ with the Riemman tensors in \eqref{eq:def-t8t8R4}, the $t_8 t_8 R^4$ structure can be written as:
\begin{equation} \label{eq:t8-contract}
\begin{split}
    t_8 t_8 R^4 &= 12 \; \left( \tensor{R}{^{A_1 A_2 A_3 A_4} } \tensor{R}{_{A_1 A_2 A_3 A_4} }\right)^2 \\
    &+ 192 \; \tensor{R}{_{A_1} ^{A_5} _{A_3} ^{A_6} } \tensor{R}{^{A_1 A_2 A_3 A_4} } \tensor{R}{_{A_2} ^{A_7} _{A_4} ^{A_8} } \tensor{R}{_{A_5 A_7 A_6 A_8} } \\
    &- 192 \; \tensor{R}{_{A_1 A_2 A_3} ^{A_5}} \tensor{R}{^{A_1 A_2 A_3 A_4} } \tensor{R}{_{A_4} ^{A_6 A_7 A_8} } \tensor{R}{_{A_5 A_6 A_7 A_8} } \\
    &+ 24 \; \tensor{R}{_{A_1 A_2} ^{A_5 A_6}} \tensor{R}{^{A_1 A_2 A_3 A_4} } \tensor{R}{_{A_3 A_4} ^{A_7 A_8} } \tensor{R}{_{A_5 A_6 A_7 A_8} } \\
    &+ 384 \; \tensor{R}{_{A_1} ^{A_5} _{A_3} ^{A_6} } \tensor{R}{^{A_1 A_2 A_3 A_4} } \tensor{R}{_{A_2} ^{A_7} _{A_6} ^{A_8} } \tensor{R}{_{A_4 A_8 A_5 A_7} } \\
    &- 96 \; \tensor{R}{_{A_1 A_2} ^{A_5 A_6}} \tensor{R}{^{A_1 A_2 A_3 A_4} } \tensor{R}{_{A_3 A_5} ^{A_7 A_8} } \tensor{R}{_{A_4 A_6 A_7 A_8} } \, .
\end{split}
\end{equation}

Similarly, $\epsilon_{10} \epsilon_{10} R^4$ structure denotes
\begin{equation} \label{eq:def-e10e10R4}
\begin{split}
    \epsilon_{10} \epsilon_{10} R^4 =& \, \epsilon_{10}^{C_1 C_2 A_1 \cdots A_8} \, \epsilon_{10\; C_1 C_2 B_1 \cdots B_8} \, \tensor{R}{^{B_1 B_2} _{A_1 A_2}} \, \cdots \, \tensor{R}{^{B_7 B_8} _{A_7 A_8}} \\
    =& - 2 \cdot 8! \, \tensor{\delta}{^{A_1} _{ [B_1}} \cdots \, \tensor{\delta}{^{A_8} _{B_8 ]}} \tensor{R}{^{B_1 B_2} _{A_1 A_2}} \, \cdots \, \tensor{R}{^{B_7 B_8} _{A_7 A_8}} \, ,
\end{split}
\end{equation}
where, in the last line, we have used the usual identity for contracting Levi-Civita tensors (c.f. equation (A.1) in \cite{Grimm:2018ohb}). Contracting all the Kronecker delta functions with the Riemann tensors in \eqref{eq:def-e10e10R4}, the $\epsilon_{10} \epsilon_{10} R^4$ structure can be written as:
\begin{equation} \label{eq:e10-contract}
\begin{split}
    &\quad \frac{1}{8} \epsilon_{10} \epsilon_{10} R^4 = - 12 \; \left( \tensor{R}{^{A_1 A_2 A_3 A_4} } \tensor{R}{_{A_1 A_2 A_3 A_4} }\right)^2 \\
    &- 192 \; \tensor{R}{_{A_1} ^{A_5} _{A_3} ^{A_6} } \tensor{R}{^{A_1 A_2 A_3 A_4} } \tensor{R}{_{A_2} ^{A_7} _{A_4} ^{A_8} } \tensor{R}{_{A_5 A_7 A_6 A_8} } \\
    &+ 192 \; \tensor{R}{_{A_1 A_2 A_3} ^{A_5}} \tensor{R}{^{A_1 A_2 A_3 A_4} } \tensor{R}{_{A_4} ^{A_6 A_7 A_8} } \tensor{R}{_{A_5 A_6 A_7 A_8} } \\
    &- 24 \; \tensor{R}{_{A_1 A_2} ^{A_5 A_6}} \tensor{R}{^{A_1 A_2 A_3 A_4} } \tensor{R}{_{A_3 A_4} ^{A_7 A_8} } \tensor{R}{_{A_5 A_6 A_7 A_8} } \\
    &+ 384 \; \tensor{R}{_{A_1 A_2} ^{A_5 A_6}} \tensor{R}{^{A_1 A_2 A_3 A_4} } \tensor{R}{_{A_3} ^{A_7} _{A_5} ^{A_8}} \tensor{R}{_{A_4 A_8 A_6 A_7} } \\
    &+ 384 \; \tensor{R}{_{A_1} ^{A_5} _{A_3} ^{A_6} } \tensor{R}{^{A_1 A_2 A_3 A_4} } \tensor{R}{_{A_2} ^{A_7} _{A_5} ^{A_8} } \tensor{R}{_{A_4 A_7 A_6 A_8} } - 768 \; \tensor{R}{^{A_1 A_2}} \tensor{R}{_{A_1} ^{A_3} _{A_2} ^{A_4}} \tensor{R}{_{A_3} ^{A_5 A_6 A_7}} \tensor{R}{_{A_4 A_5 A_6 A_7}} \\
    &+ 384 \; \tensor{R}{^{A_1 A_2}} \tensor{R}{_{A_1} ^{A_3 A_4 A_5}} \tensor{R}{_{A_2 A_3} ^{A_6 A_7}} \tensor{R}{_{A_4 A_5 A_6 A_7}} + 96 \; \tensor{R}{_{A_1 A_2}} \tensor{R}{^{A_1 A_2}} \; \tensor{R}{_{A_3 A_4 A_5 A_6}} \tensor{R}{^{A_3 A_4 A_5 A_6}} \\
    &- 1536 \; \tensor{R}{^{A_1 A_2}} \tensor{R}{_{A_1} ^{A_3 A_4 A_5}} \tensor{R}{_{A_2} ^{A_6} _{A_4} ^{A_7}} \tensor{R}{_{A_3 A_7 A_5 A_6}} + 768 \; \tensor{R}{^{A_1 A_2}} \tensor{R}{^{A_3 A_4}} \tensor{R}{_{A_1} ^{A_5} _{A_2} ^{A_6}} \tensor{R}{_{A_3 A_5 A_4 A_6}} \\
    &- 768 \; \tensor{R}{_{A_1} ^{A_3}} \tensor{R}{^{A_1 A_2}} \tensor{R}{_{A_2} ^{A_4 A_5 A_6}} \tensor{R}{_{A_3 A_4 A_5 A_6}} - 32 \; R \; \tensor{R}{_{A_1 A_2} ^{A_5 A_6}} \tensor{R}{^{A_1 A_2 A_3 A_4}} \tensor{R}{_{A_3 A_4 A_5 A_6}} \\
    &+ 128 \; R \; \tensor{R}{_{A_1} ^{A_5} _{A_3} ^{A_6}} \tensor{R}{^{A_1 A_2 A_3 A_4}} \tensor{R}{_{A_2 A_6 A_4 A_5}} - 768 \; \tensor{R}{^{A_1 A_2}} \tensor{R}{^{A_3 A_4}} \tensor{R}{_{A_1} ^{A_5} _{A_3} ^{A_6}} \tensor{R}{_{A_2 A_5 A_4 A_6}} \\
    &- 384 \; \tensor{R}{^{A_1 A_2}} \tensor{R}{^{A_3 A_4}} \tensor{R}{_{A_1 A_3} ^{A_5 A_6}} \tensor{R}{_{A_2 A_4 A_5 A_6}} + 1536 \; \tensor{R}{_{A_1}^{A_3}} \tensor{R}{^{A_1 A_2}} \tensor{R}{^{A_4 A_5}} \tensor{R}{_{A_2 A_4 A_3 A_5}} \\
    &+ 384 \; R \; \tensor{R}{^{A_1 A_2}} \tensor{R}{_{A_1} ^{A_3 A_4 A_5}} \tensor{R}{_{A_2 A_3 A_4 A_5}} - 24 \; R^2 \; \tensor{R}{_{A_1 A_2 A_3 A_4}} \tensor{R}{^{A_1 A_2 A_3 A_4}} \\
    &- 384 \; R \; \tensor{R}{^{A_1 A_2}} \tensor{R}{^{A_3 A_4}} \tensor{R}{_{A_1 A_3 A_2 A_4}} - 192 \; \left( \tensor{R}{_{A_1 A_2}} \tensor{R}{^{A_1 A_2}} \right)^2 + 384 \; \tensor{R}{_{A_1} ^{A_3}} \tensor{R}{^{A_1 A_2}} \tensor{R}{_{A_2} ^{A_4}} \tensor{R}{_{A_3 A_4}} \\
    &- 256 \; R \; \tensor{R}{_{A_1} ^{A_3}} \tensor{R}{^{A_1 A_2}} \tensor{R}{_{A_2 A_3}} + 96 \; R^2 \; \tensor{R}{_{A_1 A_2}} \tensor{R}{^{A_1 A_2}} - 4 \; R^4 \, .
\end{split}    
\end{equation}

Evaluating on Mathematica each of the contractions in \eqref{eq:t8-contract} and \eqref{eq:e10-contract} on the AdS$_2 \times \IS^8$ ansatz in \eqref{d0-ansatz}, adding them up, and substituting $v=v_2$ and $\beta= v_2/v_1$, we get to the result reported in equation \eqref{eq:R4-typeIIA}.

As we used in section \ref{sec:mtheory-story}, both the $\hat{t}_{8} \hat{t}_{8} \hat{R}^4$ and the $\frac{1}{24} \epsilon_{11} \epsilon_{11} \hat{R}^4$ structures appearing in the M-theory effective action in equation \eqref{eq:M-th-R4} look exactly the same as the terms we just discussed. This is, once written in terms of contractions of 11d Riemman tensors, they precisely reduce to the ones in \eqref{eq:t8-contract} and \eqref{eq:e10-contract}. Evaluating these contractions on the 11d metric ansatz in \eqref{eq:11d-ansatz} with Mathematica, we recover the result in \eqref{eq:new-eq}.

\newpage
\bibliographystyle{JHEP}
\bibliography{mybib}

\end{document}